\documentclass[trackchanges,twocolumn]{aastex7}
\usepackage{amsmath}
\usepackage{ulem}
\usepackage{color}

\begin{document}

\title{Production of High Energy Neutrinos in the Coronae of Ultraluminous X-ray Sources}

\author[orcid=0000-0001-8181-7511,sname='Kawanaka']{Norita Kawanaka}
\affiliation{Department of Physics, Graduate School of Science Tokyo Metropolitan University 1-1,
Minami-Osawa, Hachioji-shi, Tokyo 192-0397}
\email[show]{noritakawanaka@gmail.com}  

\author[orcid=0000-0003-2579-7266,sname='Kimura']{Shigeo S. Kimura} 
\affiliation{Astronomical Institute, Graduate School of Science, Tohoku University, Sendai 980-8578, Japan}
\affiliation{Frontier Research Institute for Interdisciplinary Sciences, Tohoku University, Sendai 980-8578, Japan}
\email[show]{shigeo@astr.tohoku.ac.jp}

\begin{abstract}
We investigate the possibility of high-energy neutrino production in coronae associated with ultraluminous X-ray sources (ULXs) as super-Eddington accreting black holes. Adopting the disk-wind-fed corona model, 
we calculate stochastic proton acceleration and the resulting neutrino emission from the ULX corona. We first examine whether the corona can be regarded as a collisionless plasma, which is a prerequisite for efficient non-thermal particle acceleration. We find that the collisionless condition is satisfied only when the mass accretion rate is lower than ($\sim 5\dot{M}_{\rm Edd}$). In this regime, protons are accelerated up to energies of a few tens of TeV and produce neutrinos through both photomeson ($p\gamma$) and hadronuclear ($pp$) interactions. 
The importance of hadronuclear interactions increases with accretion rate owing to the enhanced coronal density, resulting in broad neutrino spectra shaped by both $p\gamma$ and $pp$ processes.  We also derive a correlation between neutrino and X-ray luminosities, ($L_{\nu,{\rm tot}} \propto L_X^{2.4}$), and estimate the neutrino flux from nearby ULXs. We show that M82 X-1 is among the most promising targets and may be detectable at the 90\% confidence level by future neutrino observatories. These results suggest that super-Eddington accreting black holes constitute a previously unexplored class of high-energy neutrino sources.
\end{abstract}

\keywords{\uat{Astrophysical black holes}{98} --- \uat{accretion}{14} --- \uat{High Energy Cosmic Radiation}{731}}


\section{Introduction}

High-energy neutrinos are unique messengers of the most extreme astrophysical phenomena.  Since their first detection \citep{2013IceCubePRL,2013Sci...342E...1I}, the observations of astrophysical neutrinos up to PeV energies by IceCube has opened a new window for multimessenger astrophysics, raising fundamental questions about their sources and production mechanisms.

Recent observational advances from IceCube have strengthened the case that non-jetted active galactic nuclei (AGNs) could be sites of high-energy neutrino production.  In particular, IceCube has reported a statistically significant excess of neutrino events from the direction of the nearby type-2 Seyfert galaxy NGC 1068, one of the intrinsically-brightest AGNs in X-rays, with a neutrino flux peaking in the $\sim 1–10~{\rm TeV}$ energy range \citep{2022Sci...378..538I}.  In this context, the disk–corona systems of accreting black holes have emerged as promising sites for efficient neutrino production \citep{2019ApJ...880...40I, 2020PhRvL.125a1101M,2024PhRvD.109j1306M,2024ApJ...974...75F,2026ApJ..1003..116Y}.  The hard X-ray continuum exhibiting a power-law spectrum extending to tens or hundreds of keV that is ubiquitously observed in luminous X-ray binaries and AGNs is widely interpreted as originating from a hot corona, hot and optically-thin plasma surrounding the accretion disk, where Compton upscattering of soft disk photons by hot electrons generates the observed X-ray spectra \citep{1991ApJ...380L..51H}.  Such coronae are formed via evaporation of an underlying disk, and inferred to have high temperatures ($k_{\rm B}T_e \sim 10^2~{\rm keV}$) and low optical depths ($\tau_e<1$).  Since such coronae can be regarded as collisionless plasmas, a fraction of protons and electrons can attain relativistic energies \citep{2012SSRv..173..557L}.  For example, an accretion flow is considered to be turbulent due to magnetorotational instability (MRI; \citealp{1991ApJ...376..214B}), which can drive stochastic particle acceleration \citep{2015PhRvL.114f1101H, 2016ApJ...822...88K, 2019MNRAS.485..163K,2021MNRAS.506.1128S,2024PhRvL.133d5202B,2024MNRAS.530.1866S}.  When accelerated protons encounter intense photon fields in the coronae, photomeson interactions can lead to charged pion production and subsequent decay into neutrinos. This mechanism naturally links coronal X-ray emission with high-energy neutrino production and has been investigated in various AGN disk–corona models.  In photomeson interactions neutral pions are also produced and decay into gamma-ray photons.  Since the AGN disk-corona region is often optically thick with respect to gamma-ray photons due to high photon density, the outgoing gamma-ray luminosity would be highly suppressed \citep{2016PhRvL.116g1101M,2022ApJ...941L..17M, 2024ApJ...972...44D}, which is consistent with the fact that NGC 1068 is not so bright in TeV gamma rays as expected from its neutrino luminosity \citep{2019ApJ...883..135A}. 
Additional neutrino excesses associated with other X-ray bright Seyfert galaxies are also reported, indicating a potentially broader class of neutrino-loud AGNs besides NGC 1068 \citep{2024PhRvL.132j1002N,2025ApJ...988..141A,2025ApJ...981..103S,2026ApJ..1000L..37A}.  

Accretion disk coronae are expected to be present not only in AGNs but also in Galactic X-ray binaries containing stellar-mass black holes, and these systems have also been proposed as potential sources of high-energy neutrinos \citep{2024ApJ...975L..35F, 2025ApJ...985..139K}.
The mass accretion rates of typical AGNs and X-ray binaries are considered to be below the Eddington value, $\dot{M}_{\rm Edd}=L_{\rm Edd}/(\eta c^2)$, where $L_{\rm Edd}$ and $\eta$ are the Eddington luminosity and the radiation efficiency, respectively.  Beyond these systems, ultraluminous X-ray sources (ULXs) and some microquasars such as GRS 1915+105 suggest the existence of stellar-mass black holes accreting at super-Eddington rates \citep{2001ApJ...549L..77W, 2001ApJ...552L.109K, 2007MNRAS.377.1187P, 2004MNRAS.349..393D, 2007A&ARv..15....1D, 2010PASJ...62..239V}.  Multi-dimensional radiation-hydrodynamic (RHD) simulations performed by \citealp{2005ApJ...628..368O} (see also \citealp{2009PASJ...61L...7O, 2011ApJ...736....2O}) have shown that super-Eddington accretion flows are characterized by high mass inflow rates, strong radiation pressure, and powerful disk winds, which can modify both the structure and thermodynamic properties of the accretion flow as well as its surroundings.  \cite{2012ApJ...752...18K} calculated the radiation spectra of super-Eddington accretion flows using Monte Carlo simulations and showed that the electrons in the outflow should Compton upscatter soft photons from the underlying disk, modifying the X-ray spectra from super-Eddington accretion flows (see also \citealp{2017PASJ...69...92K}).  Observationally, ULXs and some bright microquasars often exhibit steeper X-ray spectra than typical X-ray binaries, implying the presence of relatively cooler and optically thick coronae \citep{2009MNRAS.397.1836G, 2013PASJ...65...48Y, 2017ApJ...839...46S, 2010PASJ...62..239V}. 
Similar X-ray spectral features have been reported in super-Eddington accreting AGNs including narrow-line Seyfert 1 (NLS1) galaxies \citep{2017MNRAS.468.3489K, 2022MNRAS.509.3599T, 2023A&A...678A.201Z}, implying that disk wind–driven coronae may also be formed in these sources.

If the coronae in super-Eddington accreting black holes can be regarded as collisionless plasmas, then, as in sub-Eddington systems, non-thermal particle acceleration is expected to operate efficiently.  Such processes can lead to the production of high-energy particles and, consequently, high-energy neutrino emission via hadronic interactions.  The physical properties and formation mechanisms of coronae in super-Eddington accreting black holes differ significantly from those in sub-Eddington systems.  In particular, the coronae in super-Eddington sources are likely sustained by radiation-driven disk winds rather than by evaporation from the disk, leading to differences in density, temperature, and optical depth \citep{2021PASJ...73..630K, 2025PASJ...77..811I}.  These differences may substantially affect the efficiency of particle acceleration and the resulting neutrino production.  Although the possibility of high-energy neutrino emission from ULXs has been discussed in previous studies \citep{2025A&A...698A.188P, 2025A&A...701A..98D, 2026APh...17703214P}, particle acceleration processes within coronae specific to super-Eddington accretion flows have not yet been explored in detail\footnote{\cite{2019ApJ...886..114H} discussed neutrino emission from super-Eddington accretion flows in the context of tidal disruption events.}.

In this work, we investigate particle acceleration and the associated high-energy neutrino emission in ULX coronae based on the corona model developed by \cite{2021PASJ...73..630K},  
They take into account the radiation-pressure-driven disk wind and Compton upscattering of soft photons therein.  According to this model, the corona that is fed by the disk wind has an electron-scattering optical depth of $\tau_{\rm es}\gtrsim 1$ and a temperature of $T_{\rm cor}\lesssim 10~{\rm keV}$, fairly reproducing the observed X-ray spectral properties of ULXs and other super-Eddington accreting BHs.
Using this framework, we calculate the resulting neutrino spectra and luminosities, and further examine the correlation between neutrino luminosity and X-ray luminosity.  Based on this relation, we also discuss the detectability of neutrino emission from known ULXs with current and future neutrino observatories.

\section{Model}
\subsection{Disk-corona model for super-Eddington accreting BHs}
First we briefly describe the physical model of a corona associated with a super-Eddington accretion disk (see \citealp{2021PASJ...73..630K} for the detail of the model).  As stated above, the coronal plasma in a super-Eddington accretion flow is supposed to be fed by the disk wind driven by radiation force.  To estimate the amount of the disk wind, we assume that the local mass accretion rate of the disk is given by the power-law in radius:
\begin{equation}
  \dot{M}(r)=
  \begin{cases}
    \dot{M}_0  & \text{if $r>r_{\rm crit}$}, \\
    \dot{M}_0\left(\frac{r}{r_{\rm crit}}\right)^s & \text{if $r\le r_{\rm crit}$},
  \end{cases}
\end{equation}
where $s$ is a constant less than unity and $r_{\rm crit}$ is the critical radius that is given by \citep{2005ApJ...628..368O}
\begin{eqnarray}
r_{\rm crit}=\frac{3\dot{M}_0c^2}{4L_{\rm Edd}}r_{\rm S},
\end{eqnarray}
where $r_{\rm S}=2GM_{\rm BH}/c^2$ is the Schwarzschild radius.  Considering that the wind velocity is the same order as the local escape velocity $v_{\rm esc} = (2GM_{\rm BH}/r)^{1/2}$, the coronal density as a function of radius $n_{\rm cor}(r)$ would be
\begin{eqnarray}
n_{\rm cor}=\frac{1}{4\pi m_p v_{\rm esc}r}\frac{d\dot{M}}{dr}=\frac{s\dot{M}_0}{4\pi m_pv_{\rm esc}r^2}\left( \frac{r}{r_{\rm crit}} \right)^s.
\end{eqnarray}
The value of $s$ is uncertain. $s\sim0.3$ is often adopted for the fitting of multi-wavelength spectrum of RIAF systems, but it can range from 0.1 -- 0.9 \citep[e.g.,][]{2014MNRAS.438.2804N}. The value of $s$ is not well constrained in super-Eddington systems. Radiation-hydrodynamic simulations have shown that mass outflow rates depend on mass accretion rate; The outflow rate is modest for the system close to the Eddington rate \citep{2014ApJ...796..106J,2024PASJ...76.1015Y}. Since we focus on the system close to the Eddington rate as discussed in the following sections, we use $s=0.15$ as our reference value, although $s\sim0.5$ is indicated in higher accretion rate systems \citep[e.g.,][]{2024MNRAS.532.4826T}.

The coronal temperature should be determined by the energy equilibrium in the corona.  Here we assume that the corona is heated by the reconnection of magnetic loops emerged from the underlying disk \citep{2002ApJ...572L.173L}, and that it is cooled via Comptonization of thermal soft photons by coronal electrons.  The equation of energy balance can then be described as
\begin{eqnarray}
\frac{B^2}{4\pi}V_{\rm A}\cdot \frac{\ell_{\rm cor}}{\ell_{\rm loop}}&=&\frac{4k_{\rm B}(T_{\rm cor}-T_{\rm rad})}{m_e c^2}cU_{\rm rad}\cdot {\rm max}(\tau_{\rm cor},\tau_{\rm cor}^2)\nonumber \\
&\equiv&F_{\rm Comp}, \label{coronaenergy}
\end{eqnarray}
where $B$, $V_{\rm A}$, $\ell_{\rm cor}$, $\ell_{\rm loop}$, $F_{\rm Comp}$, $T_{\rm cor}$, $U_{\rm rad}$ ($=aT_{\rm rad}^4$), and $\tau_{\rm cor}$ ($=n_{\rm cor}\sigma_{\rm T}\ell_{\rm cor}$) are the magnetic field strength in the corona, the Alfv\'{e}n velocity, the coronal scale height, the typical length of magnetic loops, the Comptonization flux, the coronal temperature, the energy density of the soft photon field, and the scattering optical depth of the corona, respectively.  Here we assume that the speed of energy dissipation via magnetic reconnection is similar to the Alfv\'{e}n velocity. $\ell_{\rm cor}/\ell_{\rm loop}$ is the reduction factor for the heating rate: it is smaller than unity when the coronal scale height is smaller than the looplength. Here the coronal magnetic field strength is determined by the assumption of equipartition: the magnetic energy density should be a constant fraction of the energy density in the underlying disk, $u_{\rm disk}$, as
\begin{eqnarray}
    \frac{B^2}{8\pi}\approx \eta_B\cdot u_{\rm disk}=\eta_B \cdot {\rm max}\left(u_{\rm rad,disk}, u_{\rm gas,disk} \right),
\end{eqnarray}
where $\eta_B$ is hereafter assumed as a constant and $u_{\rm rad,disk}=aT_{\rm disk}^4$ and $u_{\rm gas,disk}=\dot{M}\Omega/(2\pi r \alpha)$ are the radiation and gas energy density in the disk, respectively. 
We set $\eta_B= 0.015$, which corresponds to the maximal magnetic field strength considering the available disk energies \citep{2021PASJ...73..630K}. 
In the case with the radiation pressure-dominated disk, the resulting magnetic field strength is
\begin{eqnarray}
B\simeq 5.3\times 10^6~{\rm G}\left(\frac{\eta_B}{0.015}\right)^{1/2}\left(\frac{T_{\rm disk}}{10^7~{\rm K}}\right)^2.     
\end{eqnarray}
The coronal scale height is in principle determined by the length of magnetic loops, $\ell_{\rm loop}$; however, taking into account that the distance over which photons can propagate within the escape timescale of the wind-fed corona provides a lower limit, it is given by
\begin{eqnarray}
    \ell_{\rm cor}\approx {\rm min}\left( \frac{c}{v_{\rm esc}n_{\rm cor}\sigma_{\rm T}},\ell_{\rm loop} \right),
\end{eqnarray}
where the loop length is estimated as $\ell_{\rm loop}\approx {\rm min}(10r_S, r)$ \citep{2017PASJ...69...92K, 2021PASJ...73..630K}.  The seed photon energy density $U_{\rm rad}$ is given by the maximum of the intrinsic radiation flux from the disk and the reprocessed coronal irradiation flux.  When the former is larger, the effective temperature of the seed photon field can be described as
\begin{eqnarray}
    T_{\rm rad}&=&\left( \frac{L_{\rm Edd}}{4\pi a r^2 c} \right)^{1/4}\\
    &\simeq&8.6\times 10^6~{\rm K}\left(\frac{M}{10M_{\odot}}\right)^{-1/4}\left( \frac{r}{3r_{\rm S}}\right)^{-1/2}.
\end{eqnarray}

From these assumptions, one can evaluate the properties of the wind-fed corona above a super-Eddington accretion disk.  In the next subsection we discuss the properties of the wind-fed corona as a particle accelerator, and show how to calculate the energy spectra of non-thermal particles produced in the corona, as well as neutrinos produced via photomeson interactions.

\subsection{Non-thermal particle production in the corona}
\subsubsection{Collisionless condition}
Before considering the particle acceleration in the corona, we should check if our corona can be regarded as collisionless plasma, otherwise non-thermal particles would not be produced through stochastic acceleration in turbulence. 
For the corona to be collisionless, either of infall timescale $t_{\rm fall}$, dynamical timescale $t_{\rm dyn}$, or dissipation timescale $t_{\rm diss}$ should be shorter than the Coulomb relaxation timescales for protons, $t_{C,pp}$ and $t_{C,pe}$.  
The mathematical descriptions of important timescales are shown below (see also the supplemental material of \citealp{2020PhRvL.125a1101M}):
\begin{eqnarray}
t_{\rm fall}&=&\frac{r}{\alpha v_{\rm K}}, \\
t_{\rm dyn}&=&\frac{\ell_{\rm cor}}{v_{\rm esc}}, \\
t_{\rm diss}&=&\frac{\ell_{\rm cor}}{V_{\rm A}}, \\
t_{C,pp}&=&\frac{4\sqrt{\pi}\theta_p^{3/2}}{n_{\rm cor} \sigma_{\rm T} c \ln{\Lambda}}\left(\frac{m_p}{m_e}\right)^2, \\
t_{C,pe}&=&\sqrt{\frac{\pi}{2}}\frac{(\theta_p+\theta_e)^{3/2}}{n_{\rm cor} \sigma_{\rm T} c \ln{\Lambda}}\left(\frac{m_p}{m_e}\right),
\end{eqnarray}
where $\theta_i=k_{\rm B}T_i/(m_i c^2)$ ($i=p~{\rm or}~e$), $\ln{\Lambda}\sim 20$ is the Coulomb logarithm, and we hereafter consider a proton-electron coronal plasma.

To evaluate the Coulomb relaxation timescales related to protons, we need to derive the proton temperature.  Protons in the corona are heated via magnetic reconnection, and cooled via Coulomb interactions with electrons.  The energy equations for protons and electrons in the corona are
\begin{eqnarray}
\frac{B^2}{4\pi}V_{\rm A}\cdot \frac{\ell_{\rm cor}}{\ell_{\rm loop}}(1-\delta_e)&=&\frac{U_{{\rm th},p}}{t_{{\rm C},pe}}\ell_{\rm cor}+\frac{U_{{\rm th},p}}{t_{\rm dyn}}\ell_{\rm cor}, \label{protonenergy}\\
\frac{B^2}{4\pi}V_{\rm A}\cdot \frac{\ell_{\rm cor}}{\ell_{\rm loop}}\delta_e+\frac{U_{{\rm th},p}}{t_{{\rm C},pe}}\ell_{\rm cor}&=&F_{\rm Comp}+\frac{U_{{\rm th},e}}{t_{\rm dyn}}\ell_{\rm cor}, \label{electronenergy}
\end{eqnarray}
respectively, where $\delta_e$ is the fraction of the magnetic reconnection heating that goes into electrons.  Adding Eq.(\ref{protonenergy}) and Eq.(\ref{electronenergy}), we obtain
\begin{eqnarray}
    \frac{B^2}{4\pi}V_{\rm A}\cdot \frac{\ell_{\rm cor}}{\ell_{\rm loop}}= \frac{U_{\rm th,tot}}{t_{\rm dyn}}\ell_{\rm cor}+F_{\rm Comp}\approx F_{\rm Comp},
\end{eqnarray}
which is identical to the original energy equation of the corona, Eq.(\ref{coronaenergy}).  As for $\delta_e$, we use the prescription of \cite{2018MNRAS.478.5209C}, who gives the fitting formula to the results of two dimensional particle-in-cell simulations of transrelativisitic magnetic reconnection by \cite{2017ApJ...850...29R}.
We approximately obtain electron and radiation temperature inside coronae by solving Eq. (\ref{coronaenergy}, whereas the proton temperature is obtained by Eq. (\ref{protonenergy}. Then, we derive various timescales at each radius.

Fig. 1 depicts the important timescales of the wind-fed corona as functions of radius for various mass accretion rates.  Here we fix the black hole mass as $10M_{\odot}$, and $t_{C,pp}$ is not displayed here because it is always longer than the other relevant timescales over the entire parameter range considered.  One can see that when $\dot{M}\lesssim 5\dot{M}_{\rm Edd}$ the magnetic dissipation timescale, $t_{\rm diss}$ is shorter than the Coulomb relaxation timescales for protons (i.e., $t_{{\rm C},pp}$ and $t_{{\rm C},pe}$), which means that the corona can be regarded as a collisonless plasma.

\begin{figure*}[t]
    \begin{tabular}{ccc}
        \begin{minipage}{.3\textwidth}
            \centering
            \includegraphics[width=0.7\linewidth, angle=-90]{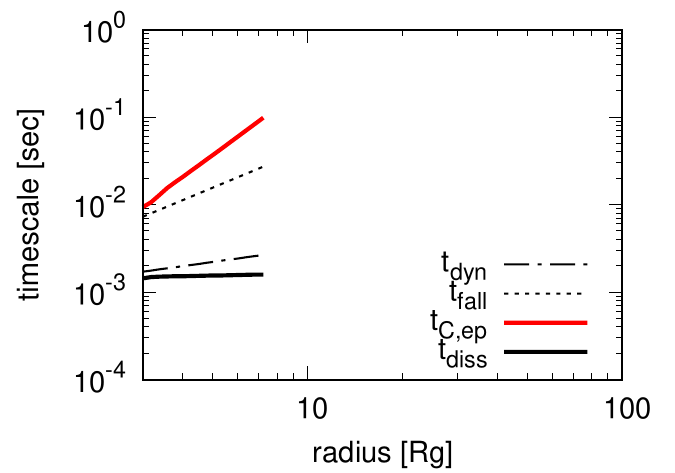}
        \end{minipage}
        \begin{minipage}{.3\textwidth}
            \centering
            \includegraphics[width=0.7\linewidth, angle=-90]{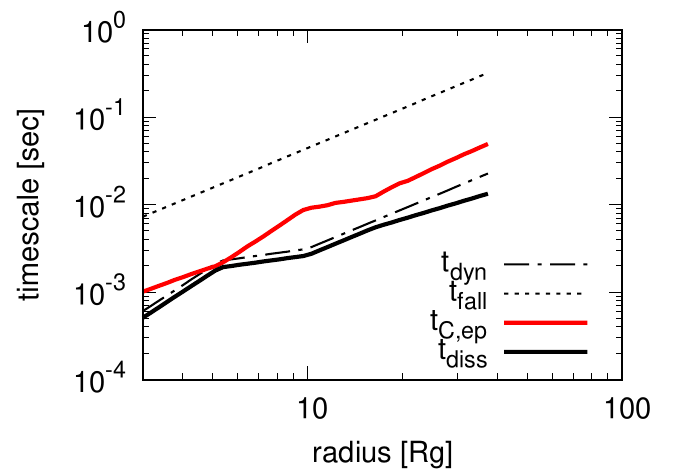}
        \end{minipage}
        \begin{minipage}{.3\textwidth}
            \centering
            \includegraphics[width=0.7\linewidth, angle=-90]{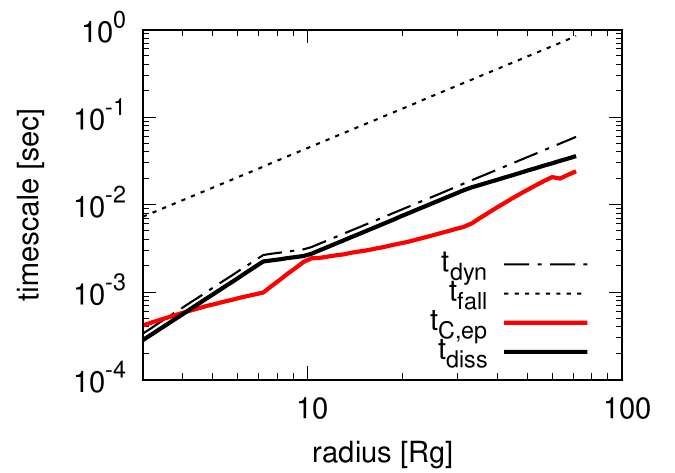}
        \end{minipage}
    \end{tabular}
\caption{Various important timescales of the corona associated with mass accretion rates of $1\dot{M}_{\rm Edd}$ (left), $5\dot{M}_{\rm Edd}$ (middle), and $10\dot{M}_{\rm Edd}$ (right) around a black hole with mass of $10M_{\odot}$.  Note that the outer radius of the region where the wind solution exists (i.e., the corona exists) is larger for larger $\dot{M}$.\label{f1}}
\end{figure*}

Hereafter, we only consider ULXs whose mass accretion rates are smaller than or equal to $5\dot{M}_{\rm Edd}$.
This critical Eddington ratio above which the corona is no longer collisionless, $\dot{m}_{\rm crit}=\dot{M}_{\rm crit}/\dot{M}_{\rm Edd}\simeq5$, is nearly independent of the black hole mass. The coronal density scale with $n_{\rm cor}\propto \dot{m}/M_{\rm BH}$. This leads to $t_{C,pe}\propto M_{\rm BH}\dot{m}^{-1}$, where $\theta_p$ and $\theta_e$ do not have strong dependence on $M_{\rm BH}$ and $\dot{m}$. On the other hand, we can write $t_{\rm diss}\propto H_{\rm cor}\propto M_{\rm BH}$, where we use $V_A$ is independent of $M_{\rm BH}$ and $\dot{m}$. Equating $t_{\rm diss}$ and $t_{C,pe}$, we obtain $\dot{m}\propto M_{\rm BH}^0$.

\subsection{Reduction to a One-zone Model}
In the original model of \cite{2021PASJ...73..630K}, the physical properties characterizing the corona -- namely the density $n_{\rm cor}$, temperature $T_{\rm cor}$, scale height $H_{\rm cor}$, and energy density of the soft photon field $U_{\rm rad}$ -- are given as functions of $r$.  In order to model particle acceleration and high energy neutrino production in the corona in a tractable manner, we reduce this corona model into a one-zone model in the following way.

We define representative values of the physical properties by evaluating their volume averages weighted by the local heating rate per unit volume.  Assuming the axisymmetry, the coronal volume element is given by $dV=2\pi r H_{\rm cor}(r)dr$, and for a physical quantity $X(r)$, the corresponding one-zone value is defined as
\begin{eqnarray}
    \langle X \rangle = \frac{\int_{r_{\rm in}}^{r_{\rm out}} X(r) q^+(r)2\pi r H_{\rm cor}(r) dr}{\int_{r_{\rm in}}^{r_{\rm out}} q^+(r)2\pi r H_{\rm cor}(r) dr},
\end{eqnarray}
where $q^+(r)=B^2/(4\pi)\cdot V_{\rm R}\cdot (\ell_{\rm cor}/\ell_{\rm loop})\cdot (1-\delta_e)$ is the proton heating rate via magnetic reconnection per unit volume.  This weighting scheme reflects the fact that regions with stronger energy dissipation are expected to dominate the acceleration of non-thermal protons and the resulting high-energy emission.  We assume that the target photon field for the $p\gamma$ interactions follows a Planck distribution with a characteristic temperature given by the average radiation temperature of the disk, $\langle T_{\rm rad} \rangle$. In most cases, this temperature is determined predominantly by the reprocessed radiation from the disk illuminated by coronal emission.

In the following calculations, we adopt the averaged values for various coronal properties (such as $n_{\rm cor}$, $T_{\rm cor}$, $H_{\rm cor}$ $B$, and so on) to characterize a homogeneous one-zone corona, and compute particle acceleration as well as neutrino production based on this simplified description.
\begin{table*}
\centering
\caption{Representative physical quantities of the one-zone corona model adopted in this work.}
\label{tab:corona_param}
\begin{tabular}{lccccc}
\hline
($M_{\rm BH}, \dot{M}$) &
$B$ &
$\theta_p$ &
$\theta_e$ &
$n_{\rm cor}$ &
$H_{\rm cor}$ \\
&
(G) &
&
&
(cm$^{-3}$) &
(cm) \\
\hline
($10\,M_\odot,\ \dot{M}_{\rm Edd}$)
& $1.7\times10^{7}$
& 0.213
& 0.124
& $3.7\times10^{16}$
& $3.0\times10^{7}$ \\

($10\,M_\odot,\ 3\dot{M}_{\rm Edd}$)
& $1.4\times10^{7}$
& 0.142
& 0.0696
& $4.9\times10^{16}$
& $3.8\times10^{7}$ \\

($10\,M_\odot,\ 5\dot{M}_{\rm Edd}$)
& $1.1\times10^{7}$
& 0.109
& 0.0439
& $4.9\times10^{16}$
& $5.8\times10^{7}$ \\

($20\,M_\odot,\ 5\dot{M}_{\rm Edd}$)
& $8.0\times10^{6}$
& 0.108
& 0.0437
& $2.4\times10^{16}$
& $1.2\times10^{8}$ \\

($30\,M_\odot,\ 5\dot{M}_{\rm Edd}$)
& $8.0\times10^{6}$
& 0.108
& 0.0437
& $2.4\times10^{16}$
& $1.2\times10^{8}$ \\
\hline
\end{tabular}
\end{table*}
Table~\ref{tab:corona_param} summarizes the physical quantities of the one-zone corona model adopted in this work for several combinations of BH mass and mass accretion rate. The magnetic field strength is typically of the order of $10^{6-7}~{\rm G}$, while the coronal density reaches $n_{\rm cor}\sim10^{16}~{\rm cm}^{-3}$. As the mass accretion rate increases, both the proton and electron temperatures decrease, reflecting the enhanced cooling in the dense corona. At the same time, the coronal scale height increases owing to the stronger mass loading from radiation-driven disk winds. These representative parameters are used throughout the subsequent calculations.

\subsection{Spectra of Nonthermal Protons}
Here we briefly review the second-order Fermi acceleration of particles in turbulence within accretion disk coronae, following the formulation of \citet{2015ApJ...806..159K}.  The second-order Fermi acceleration can be described by the Fokker-Planck equation,
\begin{eqnarray}
\frac{\partial}{\partial t}\mathcal{F}(p)
&=&
\frac{1}{p^2}
\frac{\partial}{\partial p}
\left[p^2 
\left(
D_p
\frac{\partial}{\partial p}\mathcal{F}(p)
+
\frac{p}{t_{\rm{cool}}} \mathcal{F}(p)
\right)
\right] \nonumber \\
&&-
\mathcal{F}(p)\left( t^{-1}_{\rm diff}+t^{-1}_{\rm adv} \right)
+
\dot{\mathcal{F}}_{\rm inj}(p), \label{FPeq}
\end{eqnarray}
where $\mathcal{F}(p)$ is the CR distribution function ($dN_p/dE_p=4\pi E_p^2\mathcal{F}(p)/c^3$), $D_p$ is the diffusion coefficient for the momentum space, $t_{\rm cool}$ is the cooling timescale, $t_{\rm diff}$ is the diffusion timescale, $t_{\rm adv}$ is the advection timescale, and $\dot{\mathcal{F}}_{\rm inj}$ is the injection rate.  Assuming that the turbulence in the corona has a power spectrum of $P_k \propto k^{-q}$, one can describe the diffusion coefficient for the momentum space as (e.g., \citealp{1996ApJ...456..106D})
\begin{eqnarray}
D_p \simeq (m_p c)^2\left( ck_{\rm min} \right) \left( \frac{V_{\rm A}}{c} \right)^2 \zeta \left( r_L k_{\rm min}\right)^{q-2} \gamma_p^q,
\end{eqnarray}
where $k_{\rm min} \sim H_{\rm cor}^{-1}$ is the minimum wave number of the turbulence generated in the corona with scale height of $H_{\rm cor}$, $V_{\rm A}$ is the Alfv\'{e}n speed (evaluated in the one-zone model), $r_{\rm L}=m_pc^2/(eB)$, $\zeta=8\pi\int P(k)dk/B_0^2$ is the ratio of the strength of turbulent fields to that of the non-turbulent fields, and $\gamma_p$ is the Lorentz factor of a proton.  The advection timescale is given by $t_{\rm adv}=R/v_{\rm esc}$, and for isotropically turbulent magnetic fields the diffusion timescale is given by (e.g., \citealp{2008ApJ...681.1725S})
\begin{eqnarray}
 t_{\rm diff}\simeq \frac{9R}{c}\zeta \left( \frac{r_L}{R} \right)^{q-2}\gamma_p^{q-2}.
\end{eqnarray}
For the cooling timescale $t_{\rm cool}$, we take into account $pp$ and $p\gamma$ interactions, the proton synchrotron emission, and the Bethe-Heitler process:
\begin{eqnarray}
t_{\rm cool}^{-1}=t_{pp}^{-1}+t_{p\gamma}^{-1}+t_{\rm sync}^{-1}+t_{\rm BH}^{-1},
\end{eqnarray}
where $t_{pp}$, $t_{p\gamma}$, and $t_{\rm sync}$ are the cooling timescales for each process.  The $pp$ cooling rate is
\begin{eqnarray}
    t_{pp}^{-1}=n_p \sigma_{pp} c K_{pp},
\end{eqnarray}
where $K_{pp} \sim 0.5$ is the proton inelasticity of the process, and $\sigma_{pp}$ is the cross section of the process. We use the $pp$ cross-section given by \cite{2014PhRvD..90l3014K}.
The $p\gamma$ cooling rate can be described as
\begin{eqnarray}
    t_{p\gamma}^{-1}=\frac{c}{2\gamma_p^2}\int_{\bar{\varepsilon}_{\rm thr}}^{\infty}d\bar{\varepsilon}\sigma_{p\gamma}\left(\bar{\varepsilon}\right)K_{p\gamma}\left( \varepsilon \right)\varepsilon \int_{\bar{\varepsilon}/\left( 2\gamma_p \right)}^{\infty}dE_{\gamma}\frac{N_{\gamma}\left(E_{\gamma}\right)}{E_{\gamma}^2}, \label{tpgamma}
\end{eqnarray}
where $\bar{\varepsilon}$ and $E_{\gamma}$ are the photon energy in the proton rest frame and that in the laboratory frame, respectively, $N_{\gamma}(E_{\gamma})$ is the occupation number of the target photon field, and $\bar{\varepsilon}_{\rm thr}=145~{\rm MeV}$ is the threshold energy of the process, and $\sigma_{p\gamma}$ and $K_{p\gamma}$ are the cross-section and the inelasticity of of $p\gamma$ process, respectively. 
We use fitting formulae for $\sigma_{p\gamma}$ and $K_{p\gamma}$ based on GEANT4 \citep[see][]{2006PhRvD..73f3002M}.
The synchrotron cooling timescale is described as
\begin{eqnarray}
    t_{\rm sync}^{-1}=\frac{4}{3}\left( \frac{m_e}{m_p}\right)^3 \frac{c\sigma_{\rm T}U_B}{m_e c^2} \gamma_p,
\end{eqnarray}
where $U_B=B^2/(8\pi)$ is the magnetic field energy density.  The Bethe-Heitler cooling rate can be described in the same form as Eq.(\ref{tpgamma}) by replacing the cross section $\sigma_{p\gamma}$ and inelasticity $K_{p\gamma}$ with $\sigma_{\rm BH}$ and $K_{\rm BH}$, respectively, where we used the fitting formula given in \cite{1992ApJ...400..181C} and \cite{1983MNRAS.204.1269S}.  

\begin{figure*}[t]
    \begin{tabular}{ccc}
        \begin{minipage}{.3\textwidth}
            \centering
            \includegraphics[width=1.0\linewidth]{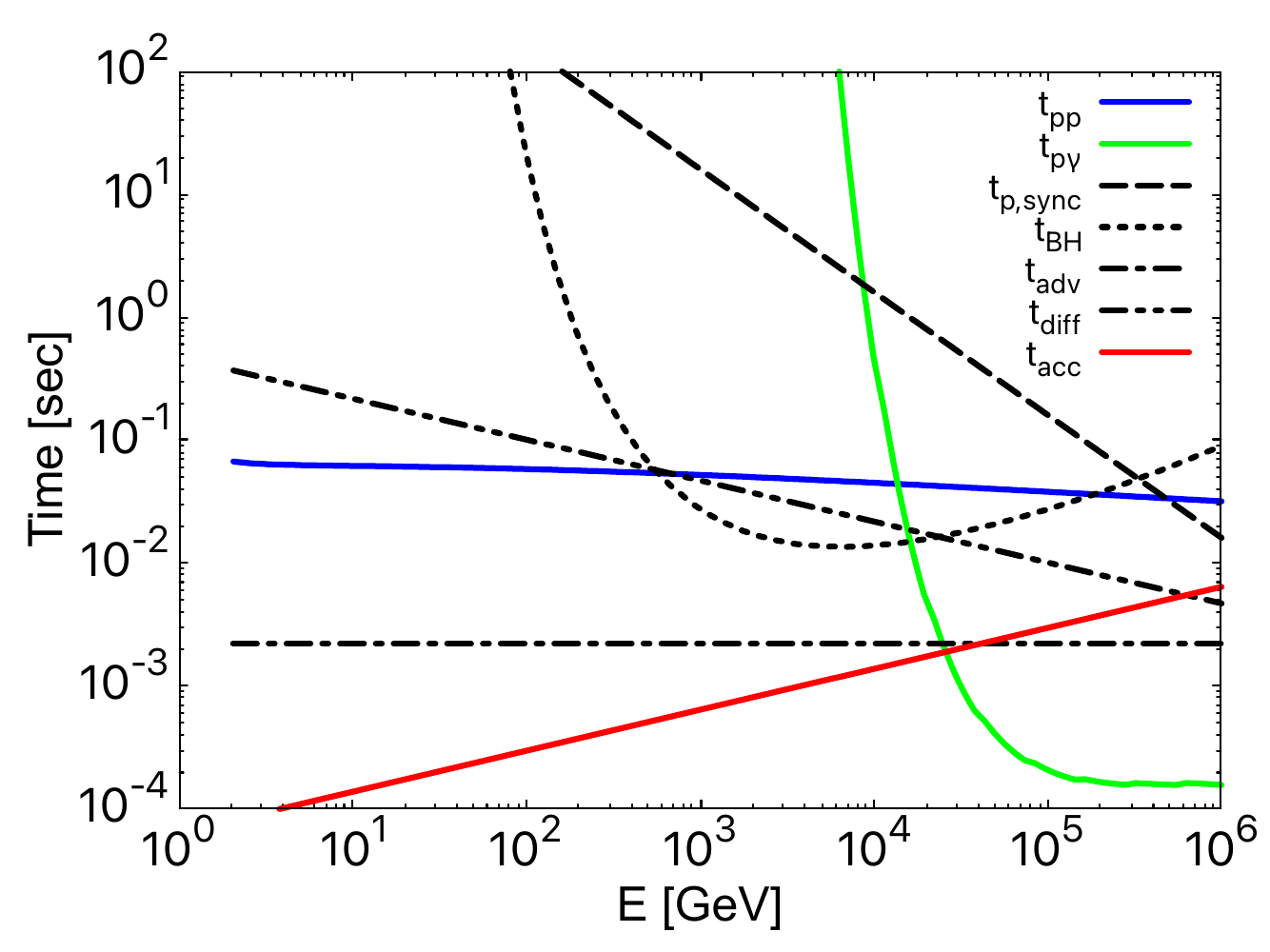}
        \end{minipage}
        \begin{minipage}{.3\textwidth}
            \centering
            \includegraphics[width=1.0\linewidth]{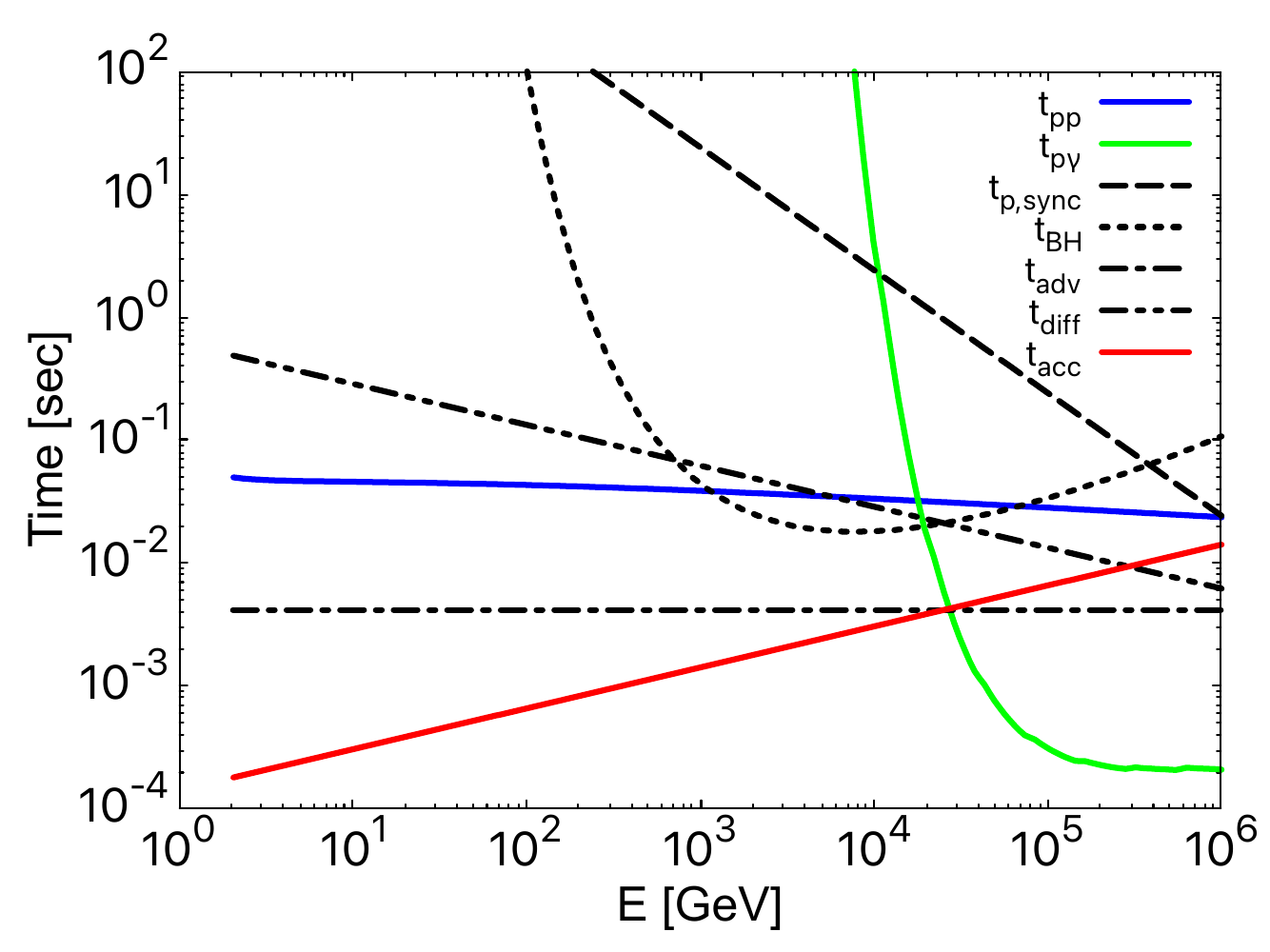}
        \end{minipage}
        \begin{minipage}{.3\textwidth}
            \centering
            \includegraphics[width=1.0\linewidth]{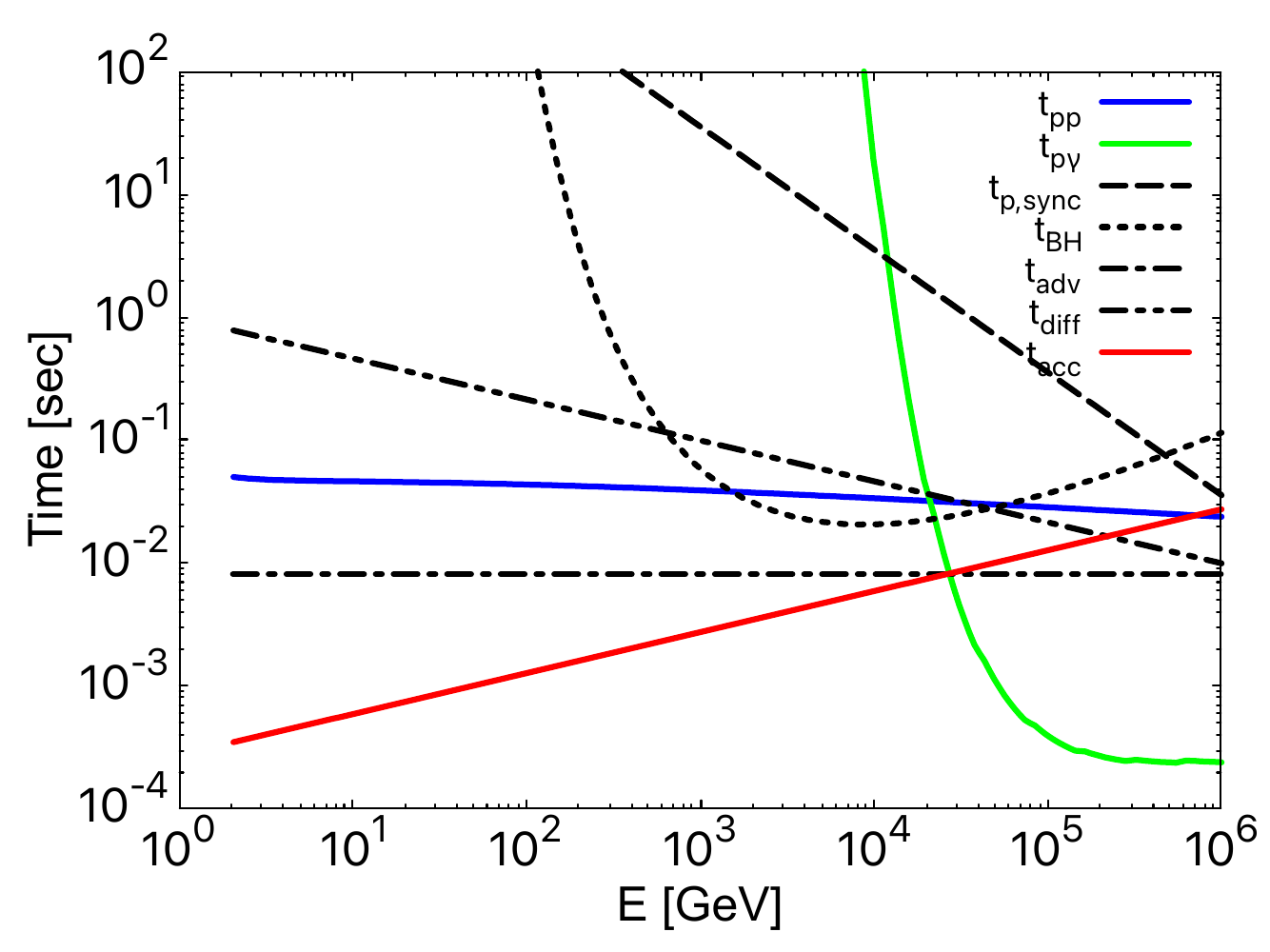}
        \end{minipage}
    \end{tabular}
\caption{Various important timescales of the corona as functions of the energy of a particle associated with mass accretion rates of $1\dot{M}_{\rm Edd}$ (left), $3\dot{M}_{\rm Edd}$ (middle), and $5\dot{M}_{\rm Edd}$ (right) around a black hole with mass of $10M_{\odot}$.\label{f2}}
\end{figure*}

Fig. \ref{f2} depicts the characteristic timescales relevant to particle acceleration and energy losses in the corona as a function of the energy of a particle for different mass accretion rates.  At proton energies below a few tens of TeV, the acceleration timescale $t_{\rm acc}$ is shorter than any of the cooling timescales, indicating that stochastic acceleration can efficiently energize particles in this energy range. At higher energies, however, $t_{p\gamma}$ becomes shorter than $t_{\rm acc}$. Therefore, photomeson interactions dominate the energy losses of accelerated protons, and the maximum proton energy is expected to be determined by the condition $t_{\rm acc}\simeq t_{p\gamma}$. Consequently, the cutoff energy of the neutrino spectrum is expected to appear at the corresponding energy scale. The present result contrasts with the situation in coronae around sub-Eddington systems in AGN, where the coronal photon field is usually dominated by disk-originated UV radiation.  In such cases, Bethe--Heitler interactions between accelerated protons and the UV photons often provide the most stringent constraint on the maximum proton energy.

As the mass accretion rate increases, the coronal density becomes higher owing to enhanced mass loading by radiation-driven disk winds. As a result, $t_{pp}$ decreases.
In this regime, a substantial fraction of accelerated protons undergoes hadronuclear interactions before escaping from the system, leading to efficient production of neutrinos through the $pp$ channel.  On the other hand, the Bethe–Heitler cooling timescale $t_{\rm BH}$ becomes shorter than $t_{pp}$ above the TeV energy range. Consequently, proton energy losses due to Bethe–Heitler pair production begin to dominate over hadronuclear interactions at high energies. This suppresses the contribution of the $pp$ channel to the neutrino spectrum at energies above $\sim {\rm TeV}$, resulting in a reduced high-energy tail of the $pp$-induced neutrino component. The importance of Bethe–Heitler cooling in shaping high-energy neutrino spectra has also been noted in previous coronal neutrino models \citep[e.g.][]{2020PhRvL.125a1101M}.

One can solve Eq.(\ref{FPeq}) until steady state is reached, assuming that the injection term is proportional to the delta function:$\dot{\mathcal{F}}_{\rm inj}=\mathcal{F}_0\delta(p-p_{\rm inj})$, where $p_{\rm inj}$ is the injection momentum and $\mathcal{F}_0$ is a constant.  Hereafter, we set $p_{\rm inj}=2m_p c$ in all calculations. The results are insensitive to this choice as long as the injection momentum is sufficiently smaller than the momentum at which the acceleration timescale is equal to the cooling timescale.  We assume that the total luminosity of relativistic protons is proportional to the
accretion luminosity, $\dot{M}c^2$, i.e.,
\begin{eqnarray}
    \eta_{\rm cr}\dot{M}c^2 = \int dV \int dp 4\pi p^2 \mathcal{F}(p)E_p\left(t_{\rm diff}^{-1}+t_{\rm fall}^{-1}\right),
\end{eqnarray}
where $\eta_{\rm cr}<1$ is the injection efficiency.  Hereafter we use $\eta_{\rm cr}=0.01$ as a fiducial value.  We solve Eq.(\ref{FPeq}) using the Chang–Cooper method \citep{1970JCoPh...6....1C}.  From the solution of the equation, one can calculate the differential neutrino luminosity. We use the fitting formulae given in \citet{2006PhRvD..74c4018K} and \citet{2008PhRvD..78c4013K} for neutrino spectra produced by $pp$ and $p\gamma$ interactions, respectively, where the distributions of neutrino energy produced by protons of energies $E_p$ are provided. One can obtain neutrino spectra by integrating the contribution of various proton energies.

We take into account suppressions of neutrino production by cooling of muons and pions by introducing suppression factors, $f_{i,\rm sup}=1-\exp(-t_{i,\rm syn}/t_{i,\rm decay})$, where $i=\pi$ or $\mu$ indicates the particle species  (see, e.g., \citealt{2023ecnp.book..433K,2025ApJ...987..218M}). The suppression by pion cooling is effective above the neutrino energy of
\begin{eqnarray}
E_{\nu,\pi,\rm sup}\approx \frac14\sqrt{\frac{6\pi m_\pi^5c^5}{m_e^2\sigma_T\tau_\pi B^2}}\simeq 3 \rm~TeV\left(\frac{B}{10^7\rm~G}\right)^{-1},
\end{eqnarray}
where $m_\pi$ and $\tau_\pi$ are the mass and the decay time of pions, respectively. Thus, we cannot expect strong neutrino signals for $E_\nu\gtrsim10$ TeV from ULXs.

\section{Results}
\begin{figure*}[t]    
    \begin{tabular}{cc}
        \begin{minipage}{.4\textwidth}
            \centering
            \includegraphics[width=0.9\linewidth]{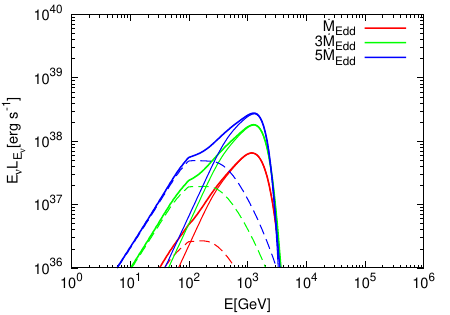}
        \end{minipage}
        \begin{minipage}{.4\textwidth}
            \centering
            \includegraphics[width=0.9\linewidth]{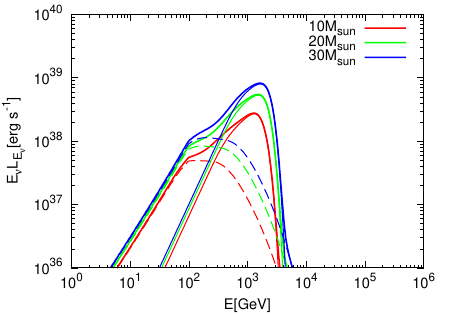}
        \end{minipage}
    \end{tabular}
\caption{High energy neutrino differential luminosities from the corona associated with super-Eddington accretion disks.  {\it Left}: spectra for different mass accretion rates ($\dot{M}=\dot{M}_{\rm Edd}, 3\dot{M}_{\rm Edd}$, and $5\dot{M}_{\rm Edd}$) with a black hole mass fixed at $10M_{\odot}$.  {\it right}: spectra for different black hole masses ($M_{\rm BH}=10M_{\odot}, 20M_{\odot}$, and $30M_{\odot}$) with the accretion rate fixed at $\dot{M}=5\dot{M}_{\rm Edd}$. Thin solid lines represent the neutrino luminosity produced via the $p\gamma$ process, the dashed lines represent that produced via the $pp$ process, and the thick solid lines show the total luminosity.\label{f3}}
\end{figure*}

The left panel of Fig.\ref{f3} depicts the neutrino spectra calculated from our model assuming a black hole mass of  $10M_{\odot}$ and mass accretion rates of $1\dot{M}_{\rm Edd}$, $3\dot{M}_{\rm Edd}$, and $5\dot{M}_{\rm Edd}$.  The spectra generally exhibit two characteristic components originating from photomeson ($p\gamma$) and hadronuclear ($pp$) interactions.  For moderate accretion rates, the neutrino spectrum is dominated by the $p\gamma$ process and exhibits a pronounced peak around 
TeV energies with a relatively sharp high-energy cutoff. The cutoff energy is determined primarily by the balance between stochastic acceleration and the cooling process.
As the accretion rate increases, the coronal density also increases because the corona is continuously supplied by radiation-pressure-driven disk winds. Consequently, proton-proton collisions become increasingly important, leading to the emergence of an additional spectral component around $E_{\nu}\sim 100~{\rm GeV}$.  The contribution of the $pp$ process becomes particularly significant for $\dot{M}\gtrsim 5\dot{M}_{\rm Edd}$, where the higher target density enhances hadronuclear interactions efficiently.

Another important feature is that the neutrino spectra become progressively broader with increasing accretion rate. This is because both $p\gamma$ and $pp$ interactions contribute simultaneously in denser coronae, producing neutrinos over a wider energy range. 
As the accretion rate increases, the enhanced coronal density increases the contribution of the $pp$ channel, resulting in broad neutrino spectra shaped by both $p\gamma$ and $pp$ processes.
The typical neutrino luminosity reaches $E_{\nu}L_{E_{\nu}}\sim 10^{38}~{\rm erg}~{\rm s}^{-1}$ around the TeV energy range for the adopted parameter set.

The right panel of Fig. \ref{f3} depicts the neutrino spectra assuming $\dot{M}=5\dot{M}_{\rm Edd}$ with $M_{\rm BH}=10$, 20, and 30 $M_\odot$. The black hole mass increases the neutrino luminosity as the released energy, $\dot{M}_c^2$ is proportional to $M_{\rm BH}$ with a fixed Eddington ratio of $\dot{m}=\dot{M}/\dot{M}_{\rm Edd}=5$. The neutrino production efficiency is always close to 1 around the peak of neutrino spectrum, $E_\nu\sim1$ TeV.

\begin{figure*}[t]    
\centering
\includegraphics{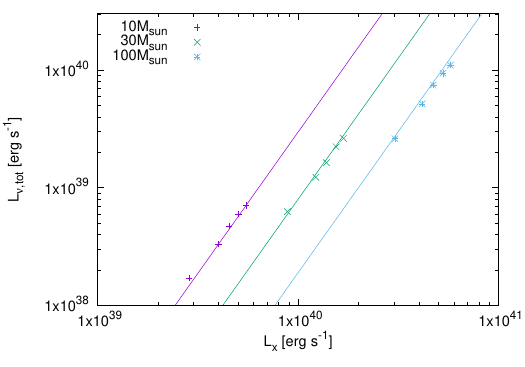}
\caption{Relations between the X-ray luminosity $L_{\rm X}$ and the integrated neutrino luminosity $L_{{\nu},{\rm tot}}$ for different black hole masses ($M_{\rm BH}=10M_{\odot}$, $30M_{\odot}$, and $100M_{\odot}$) calculated from our wind-fed corona model. The points represent the luminosities obtained for different mass accretion rates, while the solid lines show the best-fit relation, $L_{{\nu},{\rm tot}}\propto M_{\rm BH}^{-1.2}L_{\rm X}^{2.4}$.\label{f4}}
\end{figure*}

To investigate the observational implications of our model, we calculate the values of neutrino luminosity for various mass accretion rates and black-hole masses and compare them with the corresponding X-ray luminosity evaluated by the coronal model \citep{2021PASJ...73..630K}. Figure \ref{f4} shows the relation between the X-ray luminosity $L_{\rm X}$ and integrated neutrino luminosity $L_{\nu,{\rm tot}}=\int dE_{\nu} L_{E_{\nu}}$. 
We find a strong correlation approximately described by $L_{\nu,{\rm tot}}\propto L_{\rm X}^{2.4}M_{\rm BH}^{-1.2}$.  
This steep dependence of $L_{\nu,\rm tot}$ on $L_X$ originates from the weak dependence of X-ray luminosity on the accretion rate; 
We can approximately write $L_X\propto \dot{m}^{0.4}M_{\rm BH}^{0.9}$. Then, considering the neutrino production efficiency is almost unity, we can write $L_{\nu,\rm tot}\propto \dot{m}M_{\rm BH}\propto L_X^{2.5}M_{\rm BH}^{-1.3}$, which is roughly consistent with our numerical results. 

Using this relation, we estimate the detectability of known ULXs by current and future neutrino detectors. Several bright ULXs with $F_{\rm X}\gtrsim 10^{-11}~{\rm erg}~{\rm cm}^{-2}~{\rm s}^{-1}$ are expected to be promising candidates.  
However, extremely luminous systems may not necessarily be efficient neutrino emitters, because the disk wind density may increase to the point that the corona ceases to behave as a collisionless plasma for excessively high accretion rates.
In such cases, Coulomb thermalization suppresses non-thermal particle acceleration, thereby reducing the neutrino luminosity.  As discussed in Sec. 2.2., the collisionless condition for the corona is satisfied only for systems with accretion rates below approximately $\dot{M}\lesssim 5\dot{M}_{\rm Edd}$.  This condition can be used to estimate the maximum neutrino luminosity expected from a given ULX. For an observed X-ray luminosity $L_X$ the largest possible neutrino luminosity is obtained by assuming the smallest BH mass that satisfies the collisionless condition, namely the mass for which the source accretes at $\dot{M}\simeq 5\dot{M}_{\rm Edd}$.  Under this assumption, the neutrino luminosity can be regarded as an upper limit for a source with the observed X-ray luminosity.

As a representative example, we consider M82 X-1, one of the brightest ULXs known, whose X-ray luminosity is observed to be $L_X\sim 10^{41-42}~{\rm erg}~{\rm s}^{-1}$.  The nature of the compact object in M82 X-1 remains uncertain, with mass estimates ranging from several tens to several hundred solar masses \citep{2015MNRAS.451.2575S, 2016ApJ...829...28B}. 
To obtain some estimates of neutrino fluxes, we compute neutrino spectra for two parameter sets: ($M_{\rm BH},~\dot{m})=(100~M_\odot,~5)$ and ($150~M_\odot,~1.8)$.

\begin{figure*}[t]    
\centering
\includegraphics{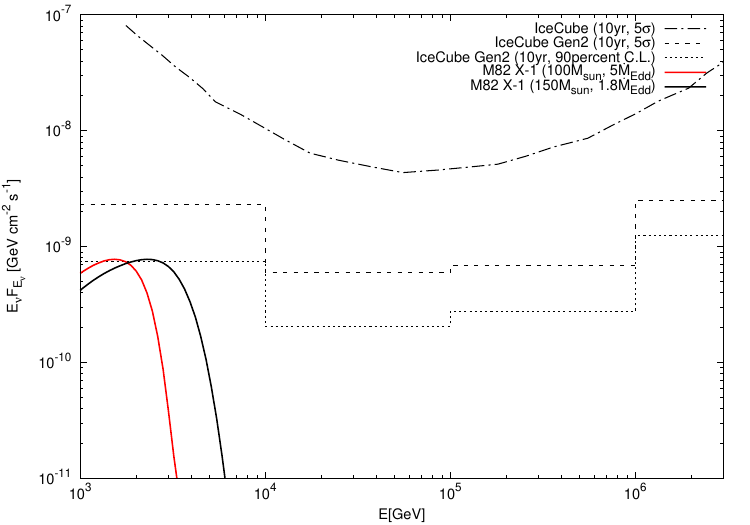}
\caption{Predicted neutrino fluxes from M82 X-1 under the assumption of a $100M_{\odot}$ black hole accreting at $\dot{M}=5\dot{M}_{\rm Edd}$ (red solid) and of a $150M_{\odot}$ black hole accreting at $\dot{M}=1.8\dot{M}_{\rm Edd}$ (black solid).  The dash-dotted and dashed curves indicate the 10-year sensitivities of IceCube and IceCube-Gen2, respectively. The predicted flux does not reach the 5$\sigma$ discovery sensitivity (thick lines) but exceeds the 90\% confidence level sensitivity (thin lines) over a limited energy range, suggesting that M82 X-1 is among the most promising ULXs for neutrino detection.\label{f5}}
\end{figure*}

Figure \ref{f5} compares the predicted neutrino fluxes with the sensitivities of IceCube and IceCube-Gen2. Although the predicted flux is insufficient to achieve a $5\sigma$ detection within a ten-year observation period, it reaches a level comparable to the 90\% confidence sensitivity of IceCube-Gen2. Consequently, M82 X-1 may represent a promising target for next-generation neutrino observation facilities. These results suggest that nearby ULXs provide a realistic opportunity to test the scenario of neutrino production in super-Eddington coronae.  Among currently known ULXs, M82 X-1 is likely to provide the most stringent observational test of the present model because of its exceptionally high X-ray flux and relatively small distance ($\sim 3.9~{\rm Mpc}$).

\section{Discussions}
\subsection{Comparison with cases with sub-Eddington accreting BHs}
Previous studies of neutrino production in accretion flows have mainly focused on radiatively inefficient accretion flows (RIAFs: \citealt{2015ApJ...806..159K}) in low-luminosity AGNs and on hot coronae associated with sub-Eddington AGNs \citep{2020PhRvL.125a1101M}. In these systems, the corona is typically characterized by a high electron temperature ($k_{\rm B}T_e\sim 100~{\rm keV}$) and a relatively small Thomson optical depth ($\tau_e \sim 1$).  

By contrast, the coronae associated with super-Eddington accretion flows exhibit substantially different physical properties: the coronae are relatively cool ($k_{\rm B}T_e\lesssim 10~{\rm keV}$) and optically thick ($\tau_e \gtrsim 5$), and for higher mass accretion rate the corona becomes more optically thick and cooler.   These differences have important consequences for high-energy neutrino production. In our model, the enhanced coronal density in higher $\dot{M}$ systems significantly increases the efficiency of $pp$ interactions, leading to the emergence of a broad neutrino component around $\sim 100~{\rm GeV}$.  

Our results suggest that ULXs may constitute a previously unexplored population of TeV neutrino sources.  The predicted neutrino luminosities, $E_{\nu}L_{E_\nu}\gtrsim 10^{39}~{\rm erg}~{\rm s}^{-1}$, are sufficiently large, and nearby systems could potentially be detected by future neutrino observatories.  The correlation obtained in this work,
$L_{\nu,{\rm tot}}\propto L_X^{2.4}M_{\rm BH}^{-1.2}$, also has important observational implications. Since the neutrino luminosity increases more rapidly than the X-ray luminosity, the brightest ULXs are expected to dominate the cumulative neutrino output from the ULX population.  However, our calculations also indicate that extremely high accretion rates may suppress efficient particle acceleration. As the mass accretion rate increases, the density of the disk-wind-fed corona rises rapidly, shortening the Coulomb relaxation timescale. Once $t_{{\rm C},ep}\lesssim t_{\rm diss}$, the plasma can no longer be regarded as collisionless, and stochastic acceleration becomes inefficient. This implies that there may exist an optimal accretion regime for neutrino production, where the target density is sufficiently high to enhance hadronic interactions while the plasma still remains collisionless.  This behavior differs from naive expectations that larger accretion rates always produce brighter neutrino emission. Instead, our results suggest the existence of a saturation or even suppression of neutrino luminosity at extremely high accretion rates.

\subsection{Model Uncertainties}
Several uncertainties remain in the present model. First, we adopted a one-zone approximation to describe the corona. In reality, the coronal density, temperature, magnetic field strength, and photon field vary significantly with radius. Although the heating-rate-weighted averaging employed here captures the dominant dissipation region, multi-zone calculations may alter the detailed neutrino spectra and luminosities.

Second, the microphysics of particle acceleration in super-Eddington coronae remains poorly understood. We assumed stochastic acceleration driven by MRI-induced turbulence, following previous studies of AGN coronae. However, magnetic reconnection may also play an important role, particularly in strongly magnetized coronal loops. The acceleration efficiency and maximum particle energy may therefore depend sensitively on the poorly constrained magnetization of the corona and magnetic field geometry.

Third, the structure of radiation-driven disk winds may affect both particle transport and neutrino escape. In particular, anisotropic outflows or clumpy winds could modify the effective density experienced by accelerated particles. Time variability of the accretion flow may also produce transient enhancements of neutrino emission.

Also, the target photon field is approximated as Planck distribution for simplicity. In reality, the photon field in the corona may have non-thermal tail due to Comptonization by $\sim10$ keV electrons, and the photon density depends on radius. To obtain accurate photon fields as a function of radius, dedicated Monte-Carlo simulations are required, which is beyond the scope of this paper.  

Finally, we have focused on stellar-mass BHs relevant to ULXs. However, similar physical processes may operate in Galactic bright microquasars such as GRS 1915+105, and in supermassive BHs accreting at super-Eddington rates, such as high-redshift quasars and some narrow-line Seyfert 1 galaxies. 
In addition, tidal disruption events and little red dots might host super-Eddington accretion flows. Most of these source classes are proposed as neutrino production sites \citep[e.g.,][]{2025A&A...698A.188P,2025A&A...701A..98D,2026APh...17703214P,2019ApJ...886..114H,2020ApJ...902..108M,2023ApJ...948...42W,2025ApJ...985..139K,2026PhRvD.113h3048K}, but the neutrino production at the coronae in these systems are not examined in detail.
Since the characteristic photon energies and dynamical timescales scale with BH mass, the resulting neutrino spectra may differ substantially from those obtained here.  Future neutrino observatories such as IceCube-Gen2, HUNT, and TRIDENT will significantly improve the sensitivity to TeV neutrino sources and may enable systematic searches for neutrino emission from ULXs, microquasars, and super-Eddington AGNs. Simultaneous X-ray observations will be particularly important because our model predicts a strong correlation between X-ray and neutrino luminosities.

\section{Summary}
In this paper, we have investigated the possibility of high-energy neutrino production in the coronae associated with super-Eddington accreting BHs. Motivated by recent observational evidence that accretion-disk coronae in Seyfert galaxies are promising sites of high-energy neutrino production, we have explored whether similar processes can operate in the distinct coronal environments of super-Eddington accretion flows.

To this end, we adopted the corona model developed by \cite{2021PASJ...73..630K}, in which the corona is supplied by radiation-driven disk winds and naturally reproduces the observed X-ray properties of ULXs and other super-Eddington sources. We calculated stochastic particle acceleration taking into account proton cooling, and the resulting neutrino emission self-consistently.

Our principal findings are summarized as follows.
\begin{enumerate}
\item
We found that the coronae of super-Eddington accreting black holes can be regarded as collisionless plasmas only when the mass accretion rate satisfies $\dot{M} \lesssim 5\dot{M}_{\rm Edd}$, and that at higher accretion rates, Coulomb relaxation becomes sufficiently efficient to suppress the formation of non-thermal particle populations. Therefore, efficient neutrino production through coronal particle acceleration is expected only in systems accreting below this critical rate.
\item
In the collisionless regime, stochastic acceleration driven by MRI-induced turbulence can accelerate protons up to energies of a few tens of TeV. The maximum proton energy is determined by the balance between particle acceleration and photomeson cooling. Consequently, the neutrino spectra exhibit cutoffs at energies corresponding to this balance.
\item
Owing to the high density of super-Eddington coronae, hadronuclear interactions become increasingly important with increasing accretion rate. In particular, when the proton-proton interaction timescale becomes shorter than the infall timescale, efficient neutrino production via the $pp$ channel is realized. On the other hand, Bethe--Heitler cooling dominates over hadronuclear interactions at higher energies, suppressing the high-energy tail of the $pp$-induced neutrino component. As a result, the predicted neutrino spectra are characterized by the coexistence of both $p\gamma$- and $pp$-induced components.
\item
We derived a strong positive correlation between the neutrino luminosity and the X-ray luminosity,
$L_{\nu,\rm tot} \propto L_X^{2.4}M_{\rm BH}^{-1.2}$, indicating that the neutrino output increases more rapidly than the X-ray luminosity. This relation enables us to estimate the neutrino luminosities of observed ULXs directly from their X-ray observations.
Applying our model to the nearby ULX M82 X-1, we showed that the expected neutrino flux may approach the sensitivity limits of future neutrino observatories. Although a $5\sigma$ detection with IceCube within a 10-year exposure appears challenging, 
a hint at the 90\% confidence level may be achieved with future facilities, such as IceCube-Gen2. Stacking analysis of bright ULXs with future facilities provide a decisive test of the present scenario.
\end{enumerate}

Our results suggest that super-Eddington accreting black holes constitute a previously unexplored population of high-energy neutrino sources. Combined X-ray and neutrino observations of nearby ULXs, microquasars, narrow-line Seyfert 1 galaxies, and other super-Eddington systems will provide valuable insights into both the origin of astrophysical neutrinos and the plasma physics of accretion-disk coronae in extreme accretion environments.

\begin{acknowledgements}
We are grateful to Y. Fujita, S. Inoue, and K. Murase for their valueable comments.  This work is supported in part by  JSPS KAKENHI Grant Number 25K01045677 (N.K.), 23H04899, 26K00733, and 26K00696 (S.S.K.).
S.S.K. acknowledges support by the Tohoku Initiative for Fostering Global Researchers for Interdisciplinary Sciences (TI-FRIS) of MEXT’s Strategic Professional Development Program for Young Researchers.
\end{acknowledgements}

\bibliography{sample7}{}

@ARTICLE{2013Sci...342E...1I,
       author = {{IceCube Collaboration}},
        title = "{Evidence for High-Energy Extraterrestrial Neutrinos at the IceCube Detector}",
      journal = {Science},
         year = 2013,
        month = nov,
       volume = {342},
       number = {6161},
          eid = {1242856},
        pages = {1242856},
          doi = {10.1126/science.1242856},
archivePrefix = {arXiv},
       eprint = {1311.5238},
 primaryClass = {astro-ph.HE},
       adsurl = {https://ui.adsabs.harvard.edu/abs/2013Sci...342E...1I}
}

@ARTICLE{2024PASJ...76.1015Y,
       author = {{Yoshioka}, Shogo and {Mineshige}, Shin and {Ohsuga}, Ken and {Kawashima}, Tomohisa and {Kitaki}, Takaaki},
        title = "{Radiation and outflow properties of super-Eddington accretion flows around various mass classes of black holes: Dependence on the accretion rates}",
      journal = {\pasj},
         year = 2024,
        month = oct,
       volume = {76},
       number = {5},
        pages = {1015-1025},
          doi = {10.1093/pasj/psae067},
archivePrefix = {arXiv},
       eprint = {2407.15927},
 primaryClass = {astro-ph.HE},
       adsurl = {https://ui.adsabs.harvard.edu/abs/2024PASJ...76.1015Y}
}

@ARTICLE{2026PhRvD.113h3048K,
       author = {{Kuze}, Riku and {Ioka}, Kunihito and {Murase}, Kohta and {Kimura}, Shigeo S. and {Inayoshi}, Kohei},
        title = "{Little red dots as hidden neutrino sources}",
      journal = {\prd},
         year = 2026,
        month = apr,
       volume = {113},
       number = {8},
          eid = {083048},
        pages = {083048},
          doi = {10.1103/vbfz-ncxd},
archivePrefix = {arXiv},
       eprint = {2601.11203},
 primaryClass = {astro-ph.HE},
       adsurl = {https://ui.adsabs.harvard.edu/abs/2026PhRvD.113h3048K}
}

@ARTICLE{2020ApJ...902..108M,
       author = {{Murase}, Kohta and {Kimura}, Shigeo S. and {Zhang}, B. Theodore and {Oikonomou}, Foteini and {Petropoulou}, Maria},
        title = "{High-energy Neutrino and Gamma-Ray Emission from Tidal Disruption Events}",
      journal = {\apj},
         year = 2020,
        month = oct,
       volume = {902},
       number = {2},
          eid = {108},
        pages = {108},
          doi = {10.3847/1538-4357/abb3c0},
archivePrefix = {arXiv},
       eprint = {2005.08937},
 primaryClass = {astro-ph.HE},
       adsurl = {https://ui.adsabs.harvard.edu/abs/2020ApJ...902..108M}
}

@ARTICLE{2023ApJ...948...42W,
       author = {{Winter}, Walter and {Lunardini}, Cecilia},
        title = "{Interpretation of the Observed Neutrino Emission from Three Tidal Disruption Events}",
      journal = {\apj},
         year = 2023,
        month = may,
       volume = {948},
       number = {1},
          eid = {42},
        pages = {42},
          doi = {10.3847/1538-4357/acbe9e},
archivePrefix = {arXiv},
       eprint = {2205.11538},
 primaryClass = {astro-ph.HE},
       adsurl = {https://ui.adsabs.harvard.edu/abs/2023ApJ...948...42W}
}

@INCOLLECTION{2023ecnp.book..433K,
       author = {{Kimura}, Shigeo S.},
        title = "{Neutrinos from Gamma-Ray Bursts}",
    booktitle = {The Encyclopedia of Cosmology. Set 2: Frontiers in Cosmology. Volume 2: Neutrino Physics and Astrophysics},
         year = 2023,
       editor = {{Stecker}, Floyd W.},
        pages = {433-482},
          doi = {10.1142/9789811282645_0009},
       adsurl = {https://ui.adsabs.harvard.edu/abs/2023ecnp.book..433K}
}

@ARTICLE{2025ApJ...987..218M,
       author = {{Mukhopadhyay}, Mainak and {Kimura}, Shigeo S. and {Metzger}, Brian D.},
        title = "{High-energy Neutrino Signatures from Pulsar Remnants of Binary Neutron-star Mergers: Coincident Detection Prospects with Gravitational Waves}",
      journal = {\apj},
         year = 2025,
        month = jul,
       volume = {987},
       number = {2},
          eid = {218},
        pages = {218},
          doi = {10.3847/1538-4357/adc913},
archivePrefix = {arXiv},
       eprint = {2407.04767},
 primaryClass = {astro-ph.HE},
       adsurl = {https://ui.adsabs.harvard.edu/abs/2025ApJ...987..218M}
}

@ARTICLE{2014ApJ...796..106J,
       author = {{Jiang}, Yan-Fei and {Stone}, James M. and {Davis}, Shane W.},
        title = "{A Global Three-dimensional Radiation Magneto-hydrodynamic Simulation of Super-Eddington Accretion Disks}",
      journal = {\apj},
         year = 2014,
        month = dec,
       volume = {796},
       number = {2},
          eid = {106},
        pages = {106},
          doi = {10.1088/0004-637X/796/2/106},
archivePrefix = {arXiv},
       eprint = {1410.0678},
 primaryClass = {astro-ph.HE},
       adsurl = {https://ui.adsabs.harvard.edu/abs/2014ApJ...796..106J}
}

@ARTICLE{2014MNRAS.438.2804N,
       author = {{Nemmen}, Rodrigo S. and {Storchi-Bergmann}, Thaisa and {Eracleous}, Michael},
        title = "{Spectral models for low-luminosity active galactic nuclei in LINERs: the role of advection-dominated accretion and jets}",
      journal = {\mnras},
         year = 2014,
        month = mar,
       volume = {438},
       number = {4},
        pages = {2804-2827},
          doi = {10.1093/mnras/stt2388},
archivePrefix = {arXiv},
       eprint = {1312.1982},
 primaryClass = {astro-ph.HE},
       adsurl = {https://ui.adsabs.harvard.edu/abs/2014MNRAS.438.2804N}
}

@ARTICLE{2016PhRvL.116g1101M,
       author = {{Murase}, Kohta and {Guetta}, Dafne and {Ahlers}, Markus},
        title = "{Hidden Cosmic-Ray Accelerators as an Origin of TeV-PeV Cosmic Neutrinos}",
      journal = {\prl},
         year = 2016,
        month = feb,
       volume = {116},
       number = {7},
          eid = {071101},
        pages = {071101},
          doi = {10.1103/PhysRevLett.116.071101},
archivePrefix = {arXiv},
       eprint = {1509.00805},
 primaryClass = {astro-ph.HE},
       adsurl = {https://ui.adsabs.harvard.edu/abs/2016PhRvL.116g1101M}
}

@ARTICLE{2024MNRAS.532.4826T,
       author = {{Toyouchi}, Daisuke and {Hotokezaka}, Kenta and {Inayoshi}, Kohei and {Kuiper}, Rolf},
        title = "{Radiation hydrodynamical simulations of super-Eddington mass transfer and black hole growth in close binaries}",
      journal = {\mnras},
         year = 2024,
        month = aug,
       volume = {532},
       number = {4},
        pages = {4826-4841},
          doi = {10.1093/mnras/stae1798},
archivePrefix = {arXiv},
       eprint = {2405.07190},
 primaryClass = {astro-ph.HE},
       adsurl = {https://ui.adsabs.harvard.edu/abs/2024MNRAS.532.4826T}
}

@ARTICLE{2024PhRvD.109j1306M,
       author = {{Mbarek}, Rostom and {Philippov}, Alexander and {Chernoglazov}, Alexander and {Levinson}, Amir and {Mushotzky}, Richard},
        title = "{Interplay between accelerated protons, x rays and neutrinos in the corona of NGC 1068: Constraints from kinetic plasma simulations}",
      journal = {\prd},
         year = 2024,
        month = may,
       volume = {109},
       number = {10},
          eid = {L101306},
        pages = {L101306},
          doi = {10.1103/PhysRevD.109.L101306},
archivePrefix = {arXiv},
       eprint = {2310.15222},
 primaryClass = {astro-ph.HE},
       adsurl = {https://ui.adsabs.harvard.edu/abs/2024PhRvD.109j1306M}
}

@ARTICLE{2024ApJ...974...75F,
       author = {{Fiorillo}, Damiano F.~G. and {Comisso}, Luca and {Peretti}, Enrico and {Petropoulou}, Maria and {Sironi}, Lorenzo},
        title = "{A Magnetized Strongly Turbulent Corona as the Source of Neutrinos from NGC 1068}",
      journal = {\apj},
         year = 2024,
        month = oct,
       volume = {974},
       number = {1},
          eid = {75},
        pages = {75},
          doi = {10.3847/1538-4357/ad7021},
archivePrefix = {arXiv},
       eprint = {2407.01678},
 primaryClass = {astro-ph.HE},
       adsurl = {https://ui.adsabs.harvard.edu/abs/2024ApJ...974...75F}
}

@ARTICLE{2026ApJ..1003..116Y,
       author = {{Yang}, Qi-Rui and {Liu}, Ruo-Yu and {Wang}, Xiang-Yu},
        title = "{Turbulent Active Galactic Nucleus Coronae as the Origin of Diffuse Neutrinos up to PeV Energies}",
      journal = {\apj},
         year = 2026,
        month = jun,
       volume = {1003},
       number = {2},
          eid = {116},
        pages = {116},
          doi = {10.3847/1538-4357/ae64f6},
archivePrefix = {arXiv},
       eprint = {2602.20969},
 primaryClass = {astro-ph.HE},
       adsurl = {https://ui.adsabs.harvard.edu/abs/2026ApJ..1003..116Y}
}

@ARTICLE{2024PhRvL.133d5202B,
       author = {{Bacchini}, Fabio and {Zhdankin}, Vladimir and {Gorbunov}, Evgeny A. and {Werner}, Gregory R. and {Arzamasskiy}, Lev and {Begelman}, Mitchell C. and {Uzdensky}, Dmitri A.},
        title = "{Collisionless Magnetorotational Turbulence in Pair Plasmas: Steady-State Dynamics, Particle Acceleration, and Radiative Cooling}",
      journal = {\prl},
         year = 2024,
        month = jul,
       volume = {133},
       number = {4},
          eid = {045202},
        pages = {045202},
          doi = {10.1103/PhysRevLett.133.045202},
archivePrefix = {arXiv},
       eprint = {2401.01399},
 primaryClass = {astro-ph.HE},
       adsurl = {https://ui.adsabs.harvard.edu/abs/2024PhRvL.133d5202B}
}

@ARTICLE{2024MNRAS.530.1866S,
       author = {{Sandoval}, Astor and {Riquelme}, Mario and {Spitkovsky}, Anatoly and {Bacchini}, Fabio},
        title = "{Particle-in-cell simulations of the magnetorotational instability in stratified shearing boxes}",
      journal = {\mnras},
         year = 2024,
        month = may,
       volume = {530},
       number = {2},
        pages = {1866-1884},
          doi = {10.1093/mnras/stae959},
archivePrefix = {arXiv},
       eprint = {2308.12348},
 primaryClass = {astro-ph.HE},
       adsurl = {https://ui.adsabs.harvard.edu/abs/2024MNRAS.530.1866S}
}

@article{2013IceCubePRL,
  title = {First Observation of PeV-Energy Neutrinos with IceCube},
  author = {Aartsen, M. G. and Abbasi, R. and Abdou, Y. and Ackermann, M. and Adams, J. and Aguilar, J. A. and Ahlers, M. and Altmann, D. and Auffenberg, J. and Bai, X. and Baker, M. and Barwick, S. W. and Baum, V. and Bay, R. and Beatty, J. J. and Bechet, S. and Becker Tjus, J. and Becker, K.-H. and Bell, M. and Benabderrahmane, M. L. and BenZvi, S. and Berdermann, J. and Berghaus, P. and Berley, D. and Bernardini, E. and Bernhard, A. and Bertrand, D. and Besson, D. Z. and Binder, G. and Bindig, D. and Bissok, M. and Blaufuss, E. and Blumenthal, J. and Boersma, D. J. and Bohaichuk, S. and Bohm, C. and Bose, D. and B\"oser, S. and Botner, O. and Brayeur, L. and Bretz, H.-P. and Brown, A. M. and Bruijn, R. and Brunner, J. and Carson, M. and Casey, J. and Casier, M. and Chirkin, D. and Christov, A. and Christy, B. and Clark, K. and Clevermann, F. and Coenders, S. and Cohen, S. and Cowen, D. F. and Cruz Silva, A. H. and Danninger, M. and Daughhetee, J. and Davis, J. C. and De Clercq, C. and De Ridder, S. and Desiati, P. and de With, M. and DeYoung, T. and D\'{\i}az-V\'elez, J. C. and Dunkman, M. and Eagan, R. and Eberhardt, B. and Eisch, J. and Ellsworth, R. W. and Euler, S. and Evenson, P. A. and Fadiran, O. and Fazely, A. R. and Fedynitch, A. and Feintzeig, J. and Feusels, T. and Filimonov, K. and Finley, C. and Fischer-Wasels, T. and Flis, S. and Franckowiak, A. and Franke, R. and Frantzen, K. and Fuchs, T. and Gaisser, T. K. and Gallagher, J. and Gerhardt, L. and Gladstone, L. and Gl\"usenkamp, T. and Goldschmidt, A. and Golup, G. and Gonzalez, J. G. and Goodman, J. A. and G\'ora, D. and Grant, D. and Gro\ss{}, A. and Gurtner, M. and Ha, C. and Haj Ismail, A. and Hallen, P. and Hallgren, A. and Halzen, F. and Hanson, K. and Heereman, D. and Heinen, D. and Helbing, K. and Hellauer, R. and Hickford, S. and Hill, G. C. and Hoffman, K. D. and Hoffmann, R. and Homeier, A. and Hoshina, K. and Huelsnitz, W. and Hulth, P. O. and Hultqvist, K. and Hussain, S. and Ishihara, A. and Jacobi, E. and Jacobsen, J. and Jagielski, K. and Japaridze, G. S. and Jero, K. and Jlelati, O. and Kaminsky, B. and Kappes, A. and Karg, T. and Karle, A. and Kelley, J. L. and Kiryluk, J. and Kislat, F. and Kl\"as, J. and Klein, S. R. and K\"ohne, J.-H. and Kohnen, G. and Kolanoski, H. and K\"opke, L. and Kopper, C. and Kopper, S. and Koskinen, D. J. and Kowalski, M. and Krasberg, M. and Krings, K. and Kroll, G. and Kunnen, J. and Kurahashi, N. and Kuwabara, T. and Labare, M. and Landsman, H. and Larson, M. J. and Lesiak-Bzdak, M. and Leuermann, M. and Leute, J. and L\"unemann, J. and Madsen, J. and Maruyama, R. and Mase, K. and Matis, H. S. and McNally, F. and Meagher, K. and Merck, M. and M\'esz\'aros, P. and Meures, T. and Miarecki, S. and Middell, E. and Milke, N. and Miller, J. and Mohrmann, L. and Montaruli, T. and Morse, R. and Nahnhauer, R. and Naumann, U. and Niederhausen, H. and Nowicki, S. C. and Nygren, D. R. and Obertacke, A. and Odrowski, S. and Olivas, A. and Olivo, M. and O'Murchadha, A. and Paul, L. and Pepper, J. A. and P\'erez de los Heros, C. and Pfendner, C. and Pieloth, D. and Pinat, E. and Pirk, N. and Posselt, J. and Price, P. B. and Przybylski, G. T. and R\"adel, L. and Rameez, M. and Rawlins, K. and Redl, P. and Reimann, R. and Resconi, E. and Rhode, W. and Ribordy, M. and Richman, M. and Riedel, B. and Rodrigues, J. P. and Rott, C. and Ruhe, T. and Ruzybayev, B. and Ryckbosch, D. and Saba, S. M. and Salameh, T. and Sander, H.-G. and Santander, M. and Sarkar, S. and Schatto, K. and Scheel, M. and Scheriau, F. and Schmidt, T. and Schmitz, M. and Schoenen, S. and Sch\"oneberg, S. and Sch\"onwald, A. and Schukraft, A. and Schulte, L. and Schulz, O. and Seckel, D. and Sestayo, Y. and Seunarine, S. and Sheremata, C. and Smith, M. W. E. and Soiron, M. and Soldin, D. and Spiczak, G. M. and Spiering, C. and Stamatikos, M. and Stanev, T. and Stasik, A. and Stezelberger, T. and Stokstad, R. G. and St\"o\ss{}l, A. and Strahler, E. A. and Str\"om, R. and Sullivan, G. W. and Taavola, H. and Taboada, I. and Tamburro, A. and Ter-Antonyan, S. and Te\ifmmode \check{s}\else \v{s}\fi{}i\ifmmode \acute{c}\else \'{c}\fi{}, G. and Tilav, S. and Toale, P. A. and Toscano, S. and Usner, M. and van der Drift, D. and van Eijndhoven, N. and Van Overloop, A. and van Santen, J. and Vehring, M. and Voge, M. and Vraeghe, M. and Walck, C. and Waldenmaier, T. and Wallraff, M. and Wasserman, R. and Weaver, Ch. and Wellons, M. and Wendt, C. and Westerhoff, S. and Whitehorn, N. and Wiebe, K. and Wiebusch, C. H. and Williams, D. R. and Wissing, H. and Wolf, M. and Wood, T. R. and Woschnagg, K. and Xu, C. and Xu, D. L. and Xu, X. W. and Yanez, J. P. and Yodh, G. and Yoshida, S. and Zarzhitsky, P. and Ziemann, J. and Zierke, S. and Zilles, A. and Zoll, M.},
  collaboration = {IceCube Collaboration},
  journal = {Phys. Rev. Lett.},
  volume = {111},
  issue = {2},
  pages = {021103},
  numpages = {7},
  year = {2013},
  month = {Jul},
  publisher = {American Physical Society},
  doi = {10.1103/PhysRevLett.111.021103},
  url = {https://link.aps.org/doi/10.1103/PhysRevLett.111.021103}
}

@ARTICLE{2006PhRvD..73f3002M,
       author = {{Murase}, Kohta and {Nagataki}, Shigehiro},
        title = "{High energy neutrino emission and neutrino background from gamma-ray bursts in the internal shock model}",
      journal = {\prd},
         year = 2006,
        month = mar,
       volume = {73},
       number = {6},
          eid = {063002},
        pages = {063002},
          doi = {10.1103/PhysRevD.73.063002},
archivePrefix = {arXiv},
       eprint = {astro-ph/0512275},
 primaryClass = {astro-ph},
       adsurl = {https://ui.adsabs.harvard.edu/abs/2006PhRvD..73f3002M}
}

@ARTICLE{2026ApJ..1000L..37A,
       author = {{Abbasi}, R. and {Ackermann}, M. and {Adams}, J. and {Agarwalla}, S.~K. and {Aguilar}, J.~A. and {Ahlers}, M. and {Alameddine}, J.~M. and {Ali}, S. and {Amin}, N.~M. and {Andeen}, K. and {Arg{\"u}elles}, C. and {Ashida}, Y. and {Athanasiadou}, S. and {Axani}, S.~N. and {Babu}, R. and {Bai}, X. and {Baines-Holmes}, J. and {Balagopal V.}, A. and {Barwick}, S.~W. and {Bash}, S. and {Basu}, V. and {Bay}, R. and {Beatty}, J.~J. and {Becker Tjus}, J. and {Behrens}, P. and {Beise}, J. and {Bellenghi}, C. and {Benkel}, S. and {BenZvi}, S. and {Berley}, D. and {Bernardini}, E. and {Besson}, D.~Z. and {Blaufuss}, E. and {Bloom}, L. and {Blot}, S. and {Bodo}, I. and {Bontempo}, F. and {Motzkin}, J.~Y. Book and {Boscolo Meneguolo}, C. and {B{\"o}ser}, S. and {Botner}, O. and {B{\"o}ttcher}, J. and {Braun}, J. and {Brinson}, B. and {Brisson-Tsavoussis}, Z. and {Burley}, R.~T. and {Butterfield}, D. and {Campana}, M.~A. and {Carloni}, K. and {Carpio}, J. and {Chattopadhyay}, S. and {Chau}, N. and {Chen}, Z. and {Chirkin}, D. and {Choi}, S. and {Clark}, B.~A. and {Coleman}, P. and {Collin}, G.~H. and {Coloma Borja}, D.~A. and {Connolly}, A. and {Conrad}, J.~M. and {Cowen}, D.~F. and {De Clercq}, C. and {DeLaunay}, J.~J. and {Delgado}, D. and {Delmeulle}, T. and {Deng}, S. and {Desiati}, P. and {de Vries}, K.~D. and {de Wasseige}, G. and {DeYoung}, T. and {D{\'\i}az-V{\'e}lez}, J.~C. and {DiKerby}, S. and {Ding}, T. and {Dittmer}, M. and {Domi}, A. and {Draper}, L. and {Dueser}, L. and {Durnford}, D. and {Dutta}, K. and {DuVernois}, M.~A. and {Ehrhardt}, T. and {Eidenschink}, L. and {Eimer}, A. and {Eldridge}, C. and {Eller}, P. and {Ellinger}, E. and {Els{\"a}sser}, D. and {Engel}, R. and {Erpenbeck}, H. and {Esmail}, W. and {Eulig}, S. and {Evans}, J. and {Evenson}, P.~A. and {Fan}, K.~L. and {Fang}, K. and {Farrag}, K. and {Fazely}, A.~R. and {Fedynitch}, A. and {Feigl}, N. and {Finley}, C. and {Fischer}, L. and {Fox}, D. and {Franckowiak}, A. and {Fukami}, S. and {F{\"u}rst}, P. and {Gallagher}, J. and {Ganster}, E. and {Garcia}, A. and {Garcia}, M. and {Garg}, G. and {Genton}, E. and {Gerhardt}, L. and {Ghadimi}, A. and {Glaser}, C. and {Gl{\"u}senkamp}, T. and {Gonzalez}, J.~G. and {Goswami}, S. and {Granados}, A. and {Grant}, D. and {Gray}, S.~J. and {Griffin}, S. and {Griswold}, S. and {Groth}, K.~M. and {Guevel}, D. and {G{\"u}nther}, C. and {Gutjahr}, P. and {Ha}, C. and {Haack}, C. and {Hallgren}, A. and {Halve}, L. and {Halzen}, F. and {Hamacher}, L. and {Ha Minh}, M. and {Handt}, M. and {Hanson}, K. and {Hardin}, J. and {Harnisch}, A.~A. and {Hatch}, P. and {Haungs}, A. and {H{\"a}ussler}, J. and {Helbing}, K. and {Hellrung}, J. and {Henke}, B. and {Hennig}, L. and {Henningsen}, F. and {Heuermann}, L. and {Hewett}, R. and {Heyer}, N. and {Hickford}, S. and {Hidvegi}, A. and {Hill}, C. and {Hill}, G.~C. and {Hmaid}, R. and {Hoffman}, K.~D. and {Hooper}, D. and {Hori}, S. and {Hoshina}, K. and {Hostert}, M. and {Hou}, W. and {Hrywniak}, M. and {Huber}, T. and {Hultqvist}, K. and {Hymon}, K. and {Ishihara}, A. and {Iwakiri}, W. and {Jacquart}, M. and {Jain}, S. and {Janik}, O. and {Jansson}, M. and {Jin}, M. and {Kamp}, N. and {Kang}, D. and {Kang}, W. and {Kappes}, A. and {Kardum}, L. and {Karg}, T. and {Karl}, M. and {Karle}, A. and {Katil}, A. and {Kauer}, M. and {Kelley}, J.~L. and {Khanal}, M. and {Khatee Zathul}, A. and {Kheirandish}, A. and {Kimku}, H. and {Kiryluk}, J. and {Klein}, C. and {Klein}, S.~R. and {Kobayashi}, Y. and {Kochocki}, A. and {Koirala}, R. and {Kolanoski}, H. and {Kontrimas}, T. and {K{\"o}pke}, L. and {Kopper}, C. and {Koskinen}, D.~J. and {Koundal}, P. and {Kowalski}, M. and {Kozynets}, T.},
        title = "{Evidence for Neutrino Emission from X-Ray Bright Seyfert Galaxies in the Southern Hemisphere Using Enhanced Starting Track Events with IceCube}",
      journal = {\apjl},
         year = 2026,
        month = apr,
       volume = {1000},
       number = {2},
          eid = {L37},
        pages = {L37},
          doi = {10.3847/2041-8213/ae4aac},
archivePrefix = {arXiv},
       eprint = {2602.10208},
 primaryClass = {astro-ph.HE},
       adsurl = {https://ui.adsabs.harvard.edu/abs/2026ApJ..1000L..37A}
}

@ARTICLE{2025ApJ...981..103S,
       author = {{Sommani}, Giacomo and {Franckowiak}, Anna and {Lincetto}, Massimiliano and {Dettmar}, Ralf-J{\"u}rgen},
        title = "{Two 100 TeV Neutrinos Coincident with the Seyfert Galaxy NGC 7469}",
      journal = {\apj},
         year = 2025,
        month = mar,
       volume = {981},
       number = {2},
          eid = {103},
        pages = {103},
          doi = {10.3847/1538-4357/adb031},
archivePrefix = {arXiv},
       eprint = {2403.03752},
 primaryClass = {astro-ph.HE},
       adsurl = {https://ui.adsabs.harvard.edu/abs/2025ApJ...981..103S}
}

@ARTICLE{2024PhRvL.132j1002N,
       author = {{Neronov}, A. and {Savchenko}, D. and {Semikoz}, D.~V.},
        title = "{Neutrino Signal from a Population of Seyfert Galaxies}",
      journal = {\prl},
         year = 2024,
        month = mar,
       volume = {132},
       number = {10},
          eid = {101002},
        pages = {101002},
          doi = {10.1103/PhysRevLett.132.101002},
archivePrefix = {arXiv},
       eprint = {2306.09018},
 primaryClass = {astro-ph.HE},
       adsurl = {https://ui.adsabs.harvard.edu/abs/2024PhRvL.132j1002N}
}

@ARTICLE{2024ApJ...972...44D,
       author = {{Das}, Abhishek and {Zhang}, B. Theodore and {Murase}, Kohta},
        title = "{Revealing the Production Mechanism of High-energy Neutrinos from NGC 1068}",
      journal = {\apj},
         year = 2024,
        month = sep,
       volume = {972},
       number = {1},
          eid = {44},
        pages = {44},
          doi = {10.3847/1538-4357/ad5a04},
archivePrefix = {arXiv},
       eprint = {2405.09332},
 primaryClass = {astro-ph.HE},
       adsurl = {https://ui.adsabs.harvard.edu/abs/2024ApJ...972...44D}
}

@ARTICLE{2014PhRvD..90l3014K,
       author = {{Kafexhiu}, Ervin and {Aharonian}, Felix and {Taylor}, Andrew M. and {Vila}, Gabriela S.},
        title = "{Parametrization of gamma-ray production cross sections for p p interactions in a broad proton energy range from the kinematic threshold to PeV energies}",
      journal = {\prd},
         year = 2014,
        month = dec,
       volume = {90},
       number = {12},
          eid = {123014},
        pages = {123014},
          doi = {10.1103/PhysRevD.90.123014},
archivePrefix = {arXiv},
       eprint = {1406.7369},
 primaryClass = {astro-ph.HE},
       adsurl = {https://ui.adsabs.harvard.edu/abs/2014PhRvD..90l3014K}
}

@ARTICLE{2008PhRvD..78c4013K,
       author = {{Kelner}, S.~R. and {Aharonian}, F.~A.},
        title = "{Energy spectra of gamma rays, electrons, and neutrinos produced at interactions of relativistic protons with low energy radiation}",
      journal = {\prd},
         year = 2008,
        month = aug,
       volume = {78},
       number = {3},
          eid = {034013},
        pages = {034013},
          doi = {10.1103/PhysRevD.78.034013},
archivePrefix = {arXiv},
       eprint = {0803.0688},
 primaryClass = {astro-ph},
       adsurl = {https://ui.adsabs.harvard.edu/abs/2008PhRvD..78c4013K}
}

@ARTICLE{2022Sci...378..538I,
       author = {{IceCube Collaboration} and {Abbasi}, R. and {Ackermann}, M. and {Adams}, J. and {Aguilar}, J.~A. and {Ahlers}, M. and {Ahrens}, M. and {Alameddine}, J.~M. and {Alispach}, C. and {Alves}, Jr., A.~A. and {Amin}, N.~M. and {Andeen}, K. and {Anderson}, T. and {Anton}, G. and {Arg{\"u}elles}, C. and {Ashida}, Y. and {Axani}, S. and {Bai}, X. and {Balagopal}, A.~V. and {Barbano}, V.~A. and {Barwick}, S.~W. and {Bastian}, B. and {Basu}, V. and {Baur}, S. and {Bay}, R. and {Beatty}, J.~J. and {Becker}, K.-H. and {Becker Tjus}, J. and {Bellenghi}, C. and {Benzvi}, S. and {Berley}, D. and {Bernardini}, E. and {Besson}, D.~Z. and {Binder}, G. and {Bindig}, D. and {Blaufuss}, E. and {Blot}, S. and {Boddenberg}, M. and {Bontempo}, F. and {Borowka}, J. and {B{\"o}ser}, S. and {Botner}, O. and {B{\"o}ttcher}, J. and {Bourbeau}, E. and {Bradascio}, F. and {Braun}, J. and {Brinson}, B. and {Bron}, S. and {Brostean-Kaiser}, J. and {Browne}, S. and {Burgman}, A. and {Burley}, R.~T. and {Busse}, R.~S. and {Campana}, M.~A. and {Carnie-Bronca}, E.~G. and {Chen}, C. and {Chen}, Z. and {Chirkin}, D. and {Choi}, K. and {Clark}, B.~A. and {Clark}, K. and {Classen}, L. and {Coleman}, A. and {Collin}, G.~H. and {Conrad}, J.~M. and {Coppin}, P. and {Correa}, P. and {Cowen}, D.~F. and {Cross}, R. and {Dappen}, C. and {Dave}, P. and {de Clercq}, C. and {Delaunay}, J.~J. and {Delgado L{\'o}pez}, D. and {Dembinski}, H. and {Deoskar}, K. and {Desai}, A. and {Desiati}, P. and {de Vries}, K.~D. and {de Wasseige}, G. and {de With}, M. and {Deyoung}, T. and {Diaz}, A. and {D{\'\i}az-V{\'e}lez}, J.~C. and {Dittmer}, M. and {Dujmovic}, H. and {Dunkman}, M. and {Duvernois}, M.~A. and {Dvorak}, E. and {Ehrhardt}, T. and {Eller}, P. and {Engel}, R. and {Erpenbeck}, H. and {Evans}, J. and {Evenson}, P.~A. and {Fan}, K.~L. and {Fazely}, A.~R. and {Fedynitch}, A. and {Feigl}, N. and {Fiedlschuster}, S. and {Fienberg}, A.~T. and {Filimonov}, K. and {Finley}, C. and {Fischer}, L. and {Fox}, D. and {Franckowiak}, A. and {Friedman}, E. and {Fritz}, A. and {F{\"u}rst}, P. and {Gaisser}, T.~K. and {Gallagher}, J. and {Ganster}, E. and {Garcia}, A. and {Garrappa}, S. and {Gerhardt}, L. and {Ghadimi}, A. and {Glaser}, C. and {Glauch}, T. and {Gl{\"u}senkamp}, T. and {Goldschmidt}, A. and {Gonzalez}, J.~G. and {Goswami}, S. and {Grant}, D. and {Gr{\'e}goire}, T. and {Griswold}, S. and {G{\"u}nther}, C. and {Gutjahr}, P. and {Haack}, C. and {Hallgren}, A. and {Halliday}, R. and {Halve}, L. and {Halzen}, F. and {Hanson}, M. Ha Minh K. and {Hardin}, J. and {Harnisch}, A.~A. and {Haungs}, A. and {Hebecker}, D. and {Helbing}, K. and {Henningsen}, F. and {Hettinger}, E.~C. and {Hickford}, S. and {Hignight}, J. and {Hill}, C. and {Hill}, G.~C. and {Hoffman}, K.~D. and {Hoffmann}, R. and {Hokanson-Fasig}, B. and {Hoshina}, K. and {Huang}, F. and {Huber}, M. and {Huber}, T. and {Hultqvist}, K. and {H{\"u}nnefeld}, M. and {Hussain}, R. and {Hymon}, K. and {in}, S. and {Iovine}, N. and {Ishihara}, A. and {Jansson}, M. and {Japaridze}, G.~S. and {Jeong}, M. and {Jin}, M. and {Jones}, B.~J.~P. and {Kang}, D. and {Kang}, W. and {Kang}, X. and {Kappes}, A. and {Kappesser}, D. and {Kardum}, L. and {Karg}, T. and {Karl}, M. and {Karle}, A. and {Katz}, U. and {Kauer}, M. and {Kellermann}, M. and {Kelley}, J.~L. and {Kheirandish}, A. and {Kin}, K. and {Kintscher}, T. and {Kiryluk}, J. and {Klein}, S.~R. and {Koirala}, R. and {Kolanoski}, H. and {Kontrimas}, T. and {K{\"o}pke}, L. and {Kopper}, C. and {Kopper}, S. and {Koskinen}, D.~J. and {Koundal}, P. and {Kovacevich}, M. and {Kowalski}, M. and {Kozynets}, T. and {Kun}, E. and {Kurahashi}, N. and {Lad}, N. and {Lagunas Gualda}, C. and {Lanfranchi}, J.~L. and {Larson}, M.~J. and {Lauber}, F. and {Lazar}, J.~P.},
        title = "{Evidence for neutrino emission from the nearby active galaxy NGC 1068}",
      journal = {Science},
         year = 2022,
        month = nov,
       volume = {378},
       number = {6619},
        pages = {538-543},
          doi = {10.1126/science.abg3395},
archivePrefix = {arXiv},
       eprint = {2211.09972},
 primaryClass = {astro-ph.HE},
       adsurl = {https://ui.adsabs.harvard.edu/abs/2022Sci...378..538I}
}

@ARTICLE{2019ApJ...880...40I,
       author = {{Inoue}, Yoshiyuki and {Khangulyan}, Dmitry and {Inoue}, Susumu and {Doi}, Akihiro},
        title = "{On High-energy Particles in Accretion Disk Coronae of Supermassive Black Holes: Implications for MeV Gamma-rays and High-energy Neutrinos from AGN Cores}",
      journal = {\apj},
         year = 2019,
        month = jul,
       volume = {880},
       number = {1},
          eid = {40},
        pages = {40},
          doi = {10.3847/1538-4357/ab2715},
archivePrefix = {arXiv},
       eprint = {1904.00554},
 primaryClass = {astro-ph.HE},
       adsurl = {https://ui.adsabs.harvard.edu/abs/2019ApJ...880...40I}
}

@ARTICLE{2020PhRvL.125a1101M,
       author = {{Murase}, Kohta and {Kimura}, Shigeo S. and {M{\'e}sz{\'a}ros}, Peter},
        title = "{Hidden Cores of Active Galactic Nuclei as the Origin of Medium-Energy Neutrinos: Critical Tests with the MeV Gamma-Ray Connection}",
      journal = {\prl},
         year = 2020,
        month = jul,
       volume = {125},
       number = {1},
          eid = {011101},
        pages = {011101},
          doi = {10.1103/PhysRevLett.125.011101},
archivePrefix = {arXiv},
       eprint = {1904.04226},
 primaryClass = {astro-ph.HE},
       adsurl = {https://ui.adsabs.harvard.edu/abs/2020PhRvL.125a1101M}
}

@ARTICLE{1991ApJ...376..214B,
       author = {{Balbus}, Steven A. and {Hawley}, John F.},
        title = "{A Powerful Local Shear Instability in Weakly Magnetized Disks. I. Linear Analysis}",
      journal = {\apj},
         year = 1991,
        month = jul,
       volume = {376},
        pages = {214},
          doi = {10.1086/170270},
       adsurl = {https://ui.adsabs.harvard.edu/abs/1991ApJ...376..214B}
}

@ARTICLE{2012SSRv..173..557L,
       author = {{Lazarian}, A. and {Vlahos}, L. and {Kowal}, G. and {Yan}, H. and {Beresnyak}, A. and {de Gouveia Dal Pino}, E.~M.},
        title = "{Turbulence, Magnetic Reconnection in Turbulent Fluids and Energetic Particle Acceleration}",
      journal = {\ssr},
         year = 2012,
        month = nov,
       volume = {173},
       number = {1-4},
        pages = {557-622},
          doi = {10.1007/s11214-012-9936-7},
archivePrefix = {arXiv},
       eprint = {1211.0008},
 primaryClass = {astro-ph.SR},
       adsurl = {https://ui.adsabs.harvard.edu/abs/2012SSRv..173..557L}
}

@ARTICLE{2019ApJ...883..135A,
       author = {{Acciari}, V.~A. and {Ansoldi}, S. and {Antonelli}, L.~A. and {Arbet Engels}, A. and {Baack}, D. and {Babi{\'c}}, A. and {Banerjee}, B. and {Barres de Almeida}, U. and {Barrio}, J.~A. and {Becerra Gonz{\'a}lez}, J. and {Bednarek}, W. and {Bellizzi}, L. and {Bernardini}, E. and {Berti}, A. and {Besenrieder}, J. and {Bhattacharyya}, W. and {Bigongiari}, C. and {Biland}, A. and {Blanch}, O. and {Bonnoli}, G. and {Bo{\v{s}}njak}, {\v{Z}}. and {Busetto}, G. and {Carosi}, R. and {Ceribella}, G. and {Chai}, Y. and {Chilingaryan}, A. and {Cikota}, S. and {Colak}, S.~M. and {Colin}, U. and {Colombo}, E. and {Contreras}, J.~L. and {Cortina}, J. and {Covino}, S. and {D'Elia}, V. and {Da Vela}, P. and {Dazzi}, F. and {De Angelis}, A. and {De Lotto}, B. and {Delfino}, M. and {Delgado}, J. and {Depaoli}, D. and {Di Pierro}, F. and {Di Venere}, L. and {Do Souto Espi{\~n}eira}, E. and {Dominis Prester}, D. and {Donini}, A. and {Dorner}, D. and {Doro}, M. and {Elsaesser}, D. and {Fallah Ramazani}, V. and {Fattorini}, A. and {Ferrara}, G. and {Fidalgo}, D. and {Foffano}, L. and {Fonseca}, M.~V. and {Font}, L. and {Fruck}, C. and {Fukami}, S. and {Garc{\'\i}a L{\'o}pez}, R.~J. and {Garczarczyk}, M. and {Gasparyan}, S. and {Gaug}, M. and {Giglietto}, N. and {Giordano}, F. and {Godinovi{\'c}}, N. and {Green}, D. and {Guberman}, D. and {Hadasch}, D. and {Hahn}, A. and {Herrera}, J. and {Hoang}, J. and {Hrupec}, D. and {H{\"u}tten}, M. and {Inada}, T. and {Inoue}, S. and {Ishio}, K. and {Iwamura}, Y. and {Jouvin}, L. and {Kerszberg}, D. and {Kubo}, H. and {Kushida}, J. and {Lamastra}, A. and {Lelas}, D. and {Leone}, F. and {Lindfors}, E. and {Lombardi}, S. and {Longo}, F. and {L{\'o}pez}, M. and {L{\'o}pez-Coto}, R. and {L{\'o}pez-Oramas}, A. and {Loporchio}, S. and {Machado de Oliveira Fraga}, B. and {Maggio}, C. and {Majumdar}, P. and {Makariev}, M. and {Mallamaci}, M. and {Maneva}, G. and {Manganaro}, M. and {Mannheim}, K. and {Maraschi}, L. and {Mariotti}, M. and {Mart{\'\i}nez}, M. and {Mazin}, D. and {Mi{\'c}anovi{\'c}}, S. and {Miceli}, D. and {Minev}, M. and {Miranda}, J.~M. and {Mirzoyan}, R. and {Molina}, E. and {Moralejo}, A. and {Morcuende}, D. and {Moreno}, V. and {Moretti}, E. and {Munar-Adrover}, P. and {Neustroev}, V. and {Nigro}, C. and {Nilsson}, K. and {Ninci}, D. and {Nishijima}, K. and {Noda}, K. and {Nogu{\'e}s}, L. and {Nozaki}, S. and {Paiano}, S. and {Palacio}, J. and {Palatiello}, M. and {Paneque}, D. and {Paoletti}, R. and {Paredes}, J.~M. and {Pe{\~n}il}, P. and {Peresano}, M. and {Persic}, M. and {Prada Moroni}, P.~G. and {Prandini}, E. and {Puljak}, I. and {Rhode}, W. and {Rib{\'o}}, M. and {Rico}, J. and {Righi}, C. and {Rugliancich}, A. and {Saha}, L. and {Sahakyan}, N. and {Saito}, T. and {Sakurai}, S. and {Satalecka}, K. and {Schmidt}, K. and {Schweizer}, T. and {Sitarek}, J. and {{\v{S}}nidari{\'c}}, I. and {Sobczynska}, D. and {Somero}, A. and {Stamerra}, A. and {Strom}, D. and {Strzys}, M. and {Suda}, Y. and {Suri{\'c}}, T. and {Takahashi}, M. and {Tavecchio}, F. and {Temnikov}, P. and {Terzi{\'c}}, T. and {Teshima}, M. and {Torres-Alb{\`a}}, N. and {Tosti}, L. and {Vagelli}, V. and {van Scherpenberg}, J. and {Vanzo}, G. and {Vazquez Acosta}, M. and {Vigorito}, C.~F. and {Vitale}, V. and {Vovk}, I. and {Will}, M. and {Zari{\'c}}, D. and {MAGIC Collaboration} and {Fiore}, F. and {Feruglio}, C. and {Rephaeli}, Y.},
        title = "{Constraints on Gamma-Ray and Neutrino Emission from NGC 1068 with the MAGIC Telescopes}",
      journal = {\apj},
         year = 2019,
        month = oct,
       volume = {883},
       number = {2},
          eid = {135},
        pages = {135},
          doi = {10.3847/1538-4357/ab3a51},
archivePrefix = {arXiv},
       eprint = {1906.10954},
 primaryClass = {astro-ph.HE},
       adsurl = {https://ui.adsabs.harvard.edu/abs/2019ApJ...883..135A}
}

@ARTICLE{2025ApJ...988..141A,
       author = {{Abbasi}, R. and {Ackermann}, M. and {Adams}, J. and {Agarwalla}, S.~K. and {Aguilar}, J.~A. and {Ahlers}, M. and {Alameddine}, J.~M. and {Amin}, N.~M. and {Andeen}, K. and {Arg{\"u}elles}, C. and {Ashida}, Y. and {Athanasiadou}, S. and {Ausborm}, L. and {Axani}, S.~N. and {Bai}, X. and {Balagopal V.}, A. and {Baricevic}, M. and {Barwick}, S.~W. and {Bash}, S. and {Basu}, V. and {Bay}, R. and {Beatty}, J.~J. and {Becker Tjus}, J. and {Beise}, J. and {Bellenghi}, C. and {Benning}, C. and {BenZvi}, S. and {Berley}, D. and {Bernardini}, E. and {Besson}, D.~Z. and {Blaufuss}, E. and {Bloom}, L. and {Blot}, S. and {Bontempo}, F. and {Book Motzkin}, J.~Y. and {Boscolo Meneguolo}, C. and {B{\"o}ser}, S. and {Botner}, O. and {B{\"o}ttcher}, J. and {Braun}, J. and {Brinson}, B. and {Brostean-Kaiser}, J. and {Brusa}, L. and {Burley}, R.~T. and {Butterfield}, D. and {Campana}, M.~A. and {Caracas}, I. and {Carloni}, K. and {Carpio}, J. and {Chattopadhyay}, S. and {Chau}, N. and {Chen}, Z. and {Chirkin}, D. and {Choi}, S. and {Clark}, B.~A. and {Coleman}, A. and {Collin}, G.~H. and {Connolly}, A. and {Conrad}, J.~M. and {Coppin}, P. and {Corley}, R. and {Correa}, P. and {Cowen}, D.~F. and {Dave}, P. and {De Clercq}, C. and {DeLaunay}, J.~J. and {Delgado}, D. and {Deng}, S. and {Desai}, A. and {Desiati}, P. and {de Vries}, K.~D. and {de Wasseige}, G. and {DeYoung}, T. and {Diaz}, A. and {D{\'\i}az-V{\'e}lez}, J.~C. and {Dierichs}, P. and {Dittmer}, M. and {Domi}, A. and {Draper}, L. and {Dujmovic}, H. and {Dutta}, K. and {DuVernois}, M.~A. and {Ehrhardt}, T. and {Eidenschink}, L. and {Eimer}, A. and {Eller}, P. and {Ellinger}, E. and {El Mentawi}, S. and {Els{\"a}sser}, D. and {Engel}, R. and {Erpenbeck}, H. and {Evans}, J. and {Evenson}, P.~A. and {Fan}, K.~L. and {Fang}, K. and {Farrag}, K. and {Fazely}, A.~R. and {Fedynitch}, A. and {Feigl}, N. and {Fiedlschuster}, S. and {Finley}, C. and {Fischer}, L. and {Fox}, D. and {Franckowiak}, A. and {Fukami}, S. and {F{\"u}rst}, P. and {Gallagher}, J. and {Ganster}, E. and {Garcia}, A. and {Garcia}, M. and {Garg}, G. and {Genton}, E. and {Gerhardt}, L. and {Ghadimi}, A. and {Girard-Carillo}, C. and {Glaser}, C. and {Glauch}, T. and {Gl{\"u}senkamp}, T. and {Gonzalez}, J.~G. and {Goswami}, S. and {Granados}, A. and {Grant}, D. and {Gray}, S.~J. and {Gries}, O. and {Griffin}, S. and {Griswold}, S. and {Groth}, K.~M. and {G{\"u}nther}, C. and {Gutjahr}, P. and {Ha}, C. and {Haack}, C. and {Hallgren}, A. and {Halve}, L. and {Halzen}, F. and {Hamdaoui}, H. and {Ha Minh}, M. and {Handt}, M. and {Hanson}, K. and {Hardin}, J. and {Harnisch}, A.~A. and {Hatch}, P. and {Haungs}, A. and {H{\"a}ussler}, J. and {Helbing}, K. and {Hellrung}, J. and {Hermannsgabner}, J. and {Heuermann}, L. and {Heyer}, N. and {Hickford}, S. and {Hidvegi}, A. and {Hill}, C. and {Hill}, G.~C. and {Hoffman}, K.~D. and {Hori}, S. and {Hoshina}, K. and {Hostert}, M. and {Hou}, W. and {Huber}, T. and {Hultqvist}, K. and {H{\"u}nnefeld}, M. and {Hussain}, R. and {Hymon}, K. and {Ishihara}, A. and {Iwakiri}, W. and {Jacquart}, M. and {Janik}, O. and {Jansson}, M. and {Japaridze}, G.~S. and {Jeong}, M. and {Jin}, M. and {Jones}, B.~J.~P. and {Kamp}, N. and {Kang}, D. and {Kang}, W. and {Kang}, X. and {Kappes}, A. and {Kappesser}, D. and {Kardum}, L. and {Karg}, T. and {Karl}, M. and {Karle}, A. and {Katil}, A. and {Katz}, U. and {Kauer}, M. and {Kelley}, J.~L. and {Khanal}, M. and {Khatee Zathul}, A. and {Kheirandish}, A. and {Kiryluk}, J. and {Klein}, S.~R. and {Kochocki}, A. and {Koirala}, R. and {Kolanoski}, H. and {Kontrimas}, T. and {K{\"o}pke}, L. and {Kopper}, C. and {Koskinen}, D.~J. and {Koundal}, P. and {Kovacevich}, M. and {Kowalski}, M.},
        title = "{IceCube Search for Neutrino Emission from X-Ray Bright Seyfert Galaxies}",
      journal = {\apj},
         year = 2025,
        month = jul,
       volume = {988},
       number = {1},
          eid = {141},
        pages = {141},
          doi = {10.3847/1538-4357/addd05},
archivePrefix = {arXiv},
       eprint = {2406.07601},
 primaryClass = {astro-ph.HE},
       adsurl = {https://ui.adsabs.harvard.edu/abs/2025ApJ...988..141A}
}

@ARTICLE{2021PASJ...73..630K,
       author = {{Kawanaka}, Norita and {Mineshige}, Shin},
        title = "{What determines the unique spectra of super-Eddington accretors? Origin of optically thick and low-temperature coronae in super-Eddington accretion flows}",
      journal = {\pasj},
         year = 2021,
        month = jun,
       volume = {73},
       number = {3},
        pages = {630-638},
          doi = {10.1093/pasj/psab023},
archivePrefix = {arXiv},
       eprint = {2012.05386},
 primaryClass = {astro-ph.HE},
       adsurl = {https://ui.adsabs.harvard.edu/abs/2021PASJ...73..630K}
}

@ARTICLE{2005ApJ...628..368O,
       author = {{Ohsuga}, Ken and {Mori}, Masao and {Nakamoto}, Taishi and {Mineshige}, Shin},
        title = "{Supercritical Accretion Flows around Black Holes: Two-dimensional, Radiation Pressure-dominated Disks with Photon Trapping}",
      journal = {\apj},
         year = 2005,
        month = jul,
       volume = {628},
       number = {1},
        pages = {368-381},
          doi = {10.1086/430728},
archivePrefix = {arXiv},
       eprint = {astro-ph/0504168},
 primaryClass = {astro-ph},
       adsurl = {https://ui.adsabs.harvard.edu/abs/2005ApJ...628..368O}
}

@ARTICLE{2002ApJ...572L.173L,
       author = {{Liu}, B.~F. and {Mineshige}, S. and {Shibata}, K.},
        title = "{A Simple Model for a Magnetic Reconnection-heated Corona}",
      journal = {\apjl},
         year = 2002,
        month = jun,
       volume = {572},
       number = {2},
        pages = {L173-L176},
          doi = {10.1086/341877},
archivePrefix = {arXiv},
       eprint = {astro-ph/0205257},
 primaryClass = {astro-ph},
       adsurl = {https://ui.adsabs.harvard.edu/abs/2002ApJ...572L.173L}
}

@ARTICLE{2017PASJ...69...92K,
       author = {{Kitaki}, Takaaki and {Mineshige}, Shin and {Ohsuga}, Ken and {Kawashima}, Tomohisa},
        title = "{Theoretical modeling of Comptonized X-ray spectra of super-Eddington accretion flow: Origin of hard excess in ultraluminous X-ray sources}",
      journal = {\pasj},
         year = 2017,
        month = dec,
       volume = {69},
       number = {6},
          eid = {92},
        pages = {92},
          doi = {10.1093/pasj/psx101},
archivePrefix = {arXiv},
       eprint = {1709.01531},
 primaryClass = {astro-ph.HE},
       adsurl = {https://ui.adsabs.harvard.edu/abs/2017PASJ...69...92K}
}

@ARTICLE{2001ApJ...549L..77W,
       author = {{Watarai}, Ken-ya and {Mizuno}, Tsunefumi and {Mineshige}, Shin},
        title = "{Slim-Disk Model for Ultraluminous X-Ray Sources}",
      journal = {\apjl},
         year = 2001,
        month = mar,
       volume = {549},
       number = {1},
        pages = {L77-L80},
          doi = {10.1086/319125},
archivePrefix = {arXiv},
       eprint = {astro-ph/0011434},
 primaryClass = {astro-ph},
       adsurl = {https://ui.adsabs.harvard.edu/abs/2001ApJ...549L..77W}
}

@ARTICLE{2001ApJ...552L.109K,
       author = {{King}, A.~R. and {Davies}, M.~B. and {Ward}, M.~J. and {Fabbiano}, G. and {Elvis}, M.},
        title = "{Ultraluminous X-Ray Sources in External Galaxies}",
      journal = {\apjl},
         year = 2001,
        month = may,
       volume = {552},
       number = {2},
        pages = {L109-L112},
          doi = {10.1086/320343},
archivePrefix = {arXiv},
       eprint = {astro-ph/0104333},
 primaryClass = {astro-ph},
       adsurl = {https://ui.adsabs.harvard.edu/abs/2001ApJ...552L.109K}
}

@ARTICLE{2007MNRAS.377.1187P,
       author = {{Poutanen}, Juri and {Lipunova}, Galina and {Fabrika}, Sergei and {Butkevich}, Alexey G. and {Abolmasov}, Pavel},
        title = "{Supercritically accreting stellar mass black holes as ultraluminous X-ray sources}",
      journal = {\mnras},
         year = 2007,
        month = may,
       volume = {377},
       number = {3},
        pages = {1187-1194},
          doi = {10.1111/j.1365-2966.2007.11668.x},
archivePrefix = {arXiv},
       eprint = {astro-ph/0609274},
 primaryClass = {astro-ph},
       adsurl = {https://ui.adsabs.harvard.edu/abs/2007MNRAS.377.1187P}
}

@ARTICLE{2004MNRAS.349..393D,
       author = {{Done}, Chris and {Wardzi{\'n}ski}, Grzegorz and {Gierli{\'n}ski}, Marek},
        title = "{GRS 1915+105: the brightest Galactic black hole}",
      journal = {\mnras},
         year = 2004,
        month = apr,
       volume = {349},
       number = {2},
        pages = {393-403},
          doi = {10.1111/j.1365-2966.2004.07545.x},
archivePrefix = {arXiv},
       eprint = {astro-ph/0308536},
 primaryClass = {astro-ph},
       adsurl = {https://ui.adsabs.harvard.edu/abs/2004MNRAS.349..393D}
}

@ARTICLE{2007A&ARv..15....1D,
       author = {{Done}, Chris and {Gierli{\'n}ski}, Marek and {Kubota}, Aya},
        title = "{Modelling the behaviour of accretion flows in X-ray binaries. Everything you always wanted to know about accretion but were afraid to ask}",
      journal = {\aapr},
         year = 2007,
        month = dec,
       volume = {15},
       number = {1},
        pages = {1-66},
          doi = {10.1007/s00159-007-0006-1},
archivePrefix = {arXiv},
       eprint = {0708.0148},
 primaryClass = {astro-ph},
       adsurl = {https://ui.adsabs.harvard.edu/abs/2007A&ARv..15....1D}
}

@ARTICLE{2010PASJ...62..239V,
       author = {{Vierdayanti}, Kiki and {Mineshige}, Shin and {Ueda}, Yoshihiro},
        title = "{Probing the Peculiar Behavior of GRS 1915+105 at Near-Eddington Luminosity}",
      journal = {\pasj},
         year = 2010,
        month = apr,
       volume = {62},
        pages = {239},
          doi = {10.1093/pasj/62.2.239},
archivePrefix = {arXiv},
       eprint = {1001.3906},
 primaryClass = {astro-ph.HE},
       adsurl = {https://ui.adsabs.harvard.edu/abs/2010PASJ...62..239V}
}

@ARTICLE{2009PASJ...61L...7O,
       author = {{Ohsuga}, Ken and {Mineshige}, Shin and {Mori}, Masao and {Kato}, Yoshiaki},
        title = "{Global Radiation-Magnetohydrodynamic Simulations of Black-Hole Accretion Flow and Outflow: Unified Model of Three States}",
      journal = {\pasj},
         year = 2009,
        month = jun,
       volume = {61},
       number = {3},
        pages = {L7-L11},
          doi = {10.1093/pasj/61.3.L7},
archivePrefix = {arXiv},
       eprint = {0903.5364},
 primaryClass = {astro-ph.HE},
       adsurl = {https://ui.adsabs.harvard.edu/abs/2009PASJ...61L...7O}
}

@ARTICLE{2011ApJ...736....2O,
       author = {{Ohsuga}, Ken and {Mineshige}, Shin},
        title = "{Global Structure of Three Distinct Accretion Flows and Outflows around Black Holes from Two-dimensional Radiation-magnetohydrodynamic Simulations}",
      journal = {\apj},
         year = 2011,
        month = jul,
       volume = {736},
       number = {1},
          eid = {2},
        pages = {2},
          doi = {10.1088/0004-637X/736/1/2},
archivePrefix = {arXiv},
       eprint = {1105.5474},
 primaryClass = {astro-ph.HE},
       adsurl = {https://ui.adsabs.harvard.edu/abs/2011ApJ...736....2O}
}

@ARTICLE{2012ApJ...752...18K,
       author = {{Kawashima}, T. and {Ohsuga}, K. and {Mineshige}, S. and {Yoshida}, T. and {Heinzeller}, D. and {Matsumoto}, R.},
        title = "{Comptonized Photon Spectra of Supercritical Black Hole Accretion Flows with Application to Ultraluminous X-Ray Sources}",
      journal = {\apj},
         year = 2012,
        month = jun,
       volume = {752},
       number = {1},
          eid = {18},
        pages = {18},
          doi = {10.1088/0004-637X/752/1/18},
       adsurl = {https://ui.adsabs.harvard.edu/abs/2012ApJ...752...18K}
}

@ARTICLE{2009MNRAS.397.1836G,
       author = {{Gladstone}, Jeanette C. and {Roberts}, Timothy P. and {Done}, Chris},
        title = "{The ultraluminous state}",
      journal = {\mnras},
         year = 2009,
        month = aug,
       volume = {397},
       number = {4},
        pages = {1836-1851},
          doi = {10.1111/j.1365-2966.2009.15123.x},
archivePrefix = {arXiv},
       eprint = {0905.4076},
 primaryClass = {astro-ph.CO},
       adsurl = {https://ui.adsabs.harvard.edu/abs/2009MNRAS.397.1836G}
}

@ARTICLE{2024ApJ...975L..35F,
       author = {{Fang}, K. and {Halzen}, Francis and {Heinz}, Sebastian and {Gallagher}, John S.},
        title = "{Astroparticles from X-Ray Binary Coronae}",
      journal = {\apjl},
         year = 2024,
        month = nov,
       volume = {975},
       number = {2},
          eid = {L35},
        pages = {L35},
          doi = {10.3847/2041-8213/ad887b},
archivePrefix = {arXiv},
       eprint = {2410.02119},
 primaryClass = {astro-ph.HE},
       adsurl = {https://ui.adsabs.harvard.edu/abs/2024ApJ...975L..35F}
}

@ARTICLE{2025ApJ...985..139K,
       author = {{Kuze}, Riku and {Kimura}, Shigeo S. and {Fang}, Ke},
        title = "{Multimessenger Emission by Magnetically Arrested Disks and Relativistic Jets of Black Hole X-Ray Binaries}",
      journal = {\apj},
         year = 2025,
        month = may,
       volume = {985},
       number = {1},
          eid = {139},
        pages = {139},
          doi = {10.3847/1538-4357/adcc1b},
archivePrefix = {arXiv},
       eprint = {2501.17467},
 primaryClass = {astro-ph.HE},
       adsurl = {https://ui.adsabs.harvard.edu/abs/2025ApJ...985..139K}
}

@ARTICLE{2013PASJ...65...48Y,
       author = {{Yoshida}, Tessei and {Isobe}, Naoki and {Mineshige}, Shin and {Kubota}, Aya and {Mizuno}, Tsunefumi and {Saitou}, Kei},
        title = "{Two Power-Law States of the Ultraluminous X-Ray Source IC 342 X-1}",
      journal = {\pasj},
         year = 2013,
        month = apr,
       volume = {65},
          eid = {48},
        pages = {48},
          doi = {10.1093/pasj/65.2.48},
archivePrefix = {arXiv},
       eprint = {1212.0994},
 primaryClass = {astro-ph.HE},
       adsurl = {https://ui.adsabs.harvard.edu/abs/2013PASJ...65...48Y}
}

@ARTICLE{2017ApJ...839...46S,
       author = {{Shidatsu}, M. and {Ueda}, Y. and {Fabrika}, S.},
        title = "{NuSTAR and Swift  Observations of the Ultraluminous X-Ray Source IC 342 X-1 in 2016: Witnessing Spectral Evolution}",
      journal = {\apj},
         year = 2017,
        month = apr,
       volume = {839},
       number = {1},
          eid = {46},
        pages = {46},
          doi = {10.3847/1538-4357/aa67e7},
archivePrefix = {arXiv},
       eprint = {1703.06399},
 primaryClass = {astro-ph.HE},
       adsurl = {https://ui.adsabs.harvard.edu/abs/2017ApJ...839...46S}
}

@ARTICLE{1991ApJ...380L..51H,
       author = {{Haardt}, F. and {Maraschi}, L.},
        title = "{A Two-Phase Model for the X-Ray Emission from Seyfert Galaxies}",
      journal = {\apjl},
         year = 1991,
        month = oct,
       volume = {380},
        pages = {L51},
          doi = {10.1086/186171},
       adsurl = {https://ui.adsabs.harvard.edu/abs/1991ApJ...380L..51H}
}

@ARTICLE{2017MNRAS.468.3489K,
       author = {{Kara}, E. and {Garc{\'\i}a}, J.~A. and {Lohfink}, A. and {Fabian}, A.~C. and {Reynolds}, C.~S. and {Tombesi}, F. and {Wilkins}, D.~R.},
        title = "{The high-Eddington NLS1 Ark 564 has the coolest corona}",
      journal = {\mnras},
         year = 2017,
        month = jul,
       volume = {468},
       number = {3},
        pages = {3489-3498},
          doi = {10.1093/mnras/stx792},
archivePrefix = {arXiv},
       eprint = {1703.09815},
 primaryClass = {astro-ph.HE},
       adsurl = {https://ui.adsabs.harvard.edu/abs/2017MNRAS.468.3489K}
}

@ARTICLE{2022MNRAS.509.3599T,
       author = {{Tortosa}, Alessia and {Ricci}, Claudio and {Tombesi}, Francesco and {Ho}, Luis C. and {Du}, Pu and {Inayoshi}, Kohei and {Wang}, Jian-Min and {Shangguan}, Jinyi and {Li}, Ruancun},
        title = "{The extreme properties of the nearby hyper-Eddington accreting active galactic nucleus in IRAS 04416+1215}",
      journal = {\mnras},
         year = 2022,
        month = jan,
       volume = {509},
       number = {3},
        pages = {3599-3615},
          doi = {10.1093/mnras/stab3152},
archivePrefix = {arXiv},
       eprint = {2109.02573},
 primaryClass = {astro-ph.GA},
       adsurl = {https://ui.adsabs.harvard.edu/abs/2022MNRAS.509.3599T}
}

@ARTICLE{2023A&A...678A.201Z,
       author = {{Zappacosta}, L. and {Piconcelli}, E. and {Fiore}, F. and {Saccheo}, I. and {Valiante}, R. and {Vignali}, C. and {Vito}, F. and {Volonteri}, M. and {Bischetti}, M. and {Comastri}, A. and {Done}, C. and {Elvis}, M. and {Giallongo}, E. and {La Franca}, F. and {Lanzuisi}, G. and {Laurenti}, M. and {Miniutti}, G. and {Bongiorno}, A. and {Brusa}, M. and {Civano}, F. and {Carniani}, S. and {D'Odorico}, V. and {Feruglio}, C. and {Gallerani}, S. and {Gilli}, R. and {Grazian}, A. and {Guainazzi}, M. and {Marinucci}, A. and {Menci}, N. and {Middei}, R. and {Nicastro}, F. and {Puccetti}, S. and {Tombesi}, F. and {Tortosa}, A. and {Testa}, V. and {Vietri}, G. and {Cristiani}, S. and {Haardt}, F. and {Maiolino}, R. and {Schneider}, R. and {Tripodi}, R. and {Vallini}, L. and {Vanzella}, E.},
        title = "{HYPerluminous quasars at the Epoch of ReionizatION (HYPERION): A new regime for the X-ray nuclear properties of the first quasars}",
      journal = {\aap},
         year = 2023,
        month = oct,
       volume = {678},
          eid = {A201},
        pages = {A201},
          doi = {10.1051/0004-6361/202346795},
archivePrefix = {arXiv},
       eprint = {2305.02347},
 primaryClass = {astro-ph.GA},
       adsurl = {https://ui.adsabs.harvard.edu/abs/2023A&A...678A.201Z}
}

@ARTICLE{2025A&A...698A.188P,
       author = {{Peretti}, Enrico and {Petropoulou}, Maria and {Vasilopoulos}, Georgios and {Gabici}, Stefano},
        title = "{Particle acceleration and multi-messenger radiation from ultra-luminous X-ray sources: A new class of Galactic PeVatrons}",
      journal = {\aap},
         year = 2025,
        month = jun,
       volume = {698},
          eid = {A188},
        pages = {A188},
          doi = {10.1051/0004-6361/202452987},
archivePrefix = {arXiv},
       eprint = {2411.08762},
 primaryClass = {astro-ph.HE},
       adsurl = {https://ui.adsabs.harvard.edu/abs/2025A&A...698A.188P}
}

@ARTICLE{2025A&A...701A..98D,
       author = {{Ducci}, L. and {Perinati}, E. and {Romano}, P. and {Vercellone}, S. and {Niko{\l}ajuk}, M. and {Santangelo}, A. and {Sasaki}, M.},
        title = "{Ultra-luminous X-ray pulsars as sources of TeV neutrinos}",
      journal = {\aap},
         year = 2025,
        month = sep,
       volume = {701},
          eid = {A98},
        pages = {A98},
          doi = {10.1051/0004-6361/202555242},
archivePrefix = {arXiv},
       eprint = {2508.10487},
 primaryClass = {astro-ph.HE},
       adsurl = {https://ui.adsabs.harvard.edu/abs/2025A&A...701A..98D}
}
\bibliographystyle{aasjournalv7}



\end{document}